# Broad Sub-Continuum Resonances and the Case for Finite-Energy Sum-Rules

A. S. Deakin, V. Elias, A. H. Fariborz, and Ying Xue
Department of Applied Mathematics
University of Western Ontario
London, Ontario N6A 5B7
Canada

and

Fang Shi and T. G. Steele
Department of Physics and Engineering Physics
University of Saskatchewan
Saskatoon, Saskatchewan S7N 5C6
Canada

*Abstract*

There is a need to go beyond the narrow resonance approximation for QCD sum-rule channels which are likely to exhibit sensitivity to broad resonance structures. We first discuss how the first two Laplace sum rules are altered when one goes beyond the narrow resonance approximation to include possible subcontinuum resonances with nonzero widths. We then show that the corresponding first two finite energy sum rules are insensitive to the widths of such resonances, provided their peaks are symmetric and entirely below the continuum threshold. We also discuss the reduced sensitivity of the first two finite energy sum rules to higher dimensional condensates, and show these sum rules to be insensitive to dimension > 6 condensates containing at least one $\overline{q}q$ pair. We extract the direct single-instanton contribution to the $F_1$ sum rule for the longitudinal component of the axial-vector correlation function from the known single-instanton contribution to the lowest Laplace sum rule for the pseudoscalar channel. Finally, we demonstrate how inclusion of this instanton contribution to the finite-energy sum rule leads to both a lighter quark mass and to more phenomenologically reasonable higher-mass-resonance contributions within the pseudoscalar channel.

## I. Introduction: Nonzero Resonance Widths and QCD Laplace Sum-Rules

*Laplace Sum-Rule Methodology in the Narrow Resonance Approximation*

Hadron properties can be extracted by relating phenomenological and field-theoretical expressions for integrals over appropriately chosen current-correlation functions, integrals which we denote as QCD sum rules [1]. The phenomenological expressions are generally extracted via the narrow resonance approximation. In the narrow resonance approximation, hadronic contributions to the imaginary part of current-current correlation functions are proportional to δ-functions at the resonance mass,

$$\mathrm{Im}[\Pi^h(s)] \quad = \quad \sum_r \pi g_r \delta(s - m_r^2) + \Theta(s - s_0)\, \mathrm{Im}[\Pi^p(s)], \tag{1}$$

The summation in (1) is over all resonances r in the channel under consideration [i.e., whose quantum numbers are consistent with the choice of currents in the current correlation function] such that $m_r^2$ is less than $s_0$. Above this hadron-continuum threshold, the hadronic contribution $\Pi^h(s)$ to the correlation function is assumed to be the same as the contribution $\Pi^p(s)$ from perturbative QCD, as is evident from (1).

The hadronic sub-continuum (h) contribution to the $k^{th}$ Laplace sum rule, corresponding to the transform of the appropriate portions of (1), is defined to be

$$R_k^h(\tau) \;\equiv\; \int_0^\infty ds\, (1/\pi)\, \mathrm{Im}[\Pi^h(s) - \Pi^p(s)\, \Theta(s - s_0)]\, s^k\, e^{-s\tau} \tag{2}$$

In the narrow resonance approximation ($\Gamma \to 0$), we see from (1) that

$$\lim_{\Gamma\to 0} R_k^h(\tau) = \sum_r g_r\, m_r^{2k} \exp[-m_r^2\tau], \qquad (3)$$

an expression in which contributions from more-massive resonances are exponentially suppressed. Note from (3) that $R_1^h(\tau) \geq m_\ell^2\, R_0^h(\tau)$, where $m_\ell$ denotes the mass of the lowest-lying resonance in the channel. Consequently, $R_1^h(\tau)/R_0^h(\tau)$ is bounded from below by $m_\ell^2$. Standard QCD sum-rule methodology involves minimizing this ratio [or its field-theoretical analogue] with respect to $\tau$ in order to determine a value of $m_\ell^2$ [2]. The sum rule $R_k^h(\tau)$ corresponds to the following field-theoretical contribution from perturbative-QCD and nonperturbative (np) QCD-vacuum effects:

$$R_k^{QCD}(\tau) = \int_0^{s_0} ds\, (1/\pi)\, \mathrm{Im}[\Pi^p(s)]\, s^k\, e^{-s\tau}$$

$$+ \; (-\partial/\partial\tau)^k \{(1/\tau)\, \mathcal{L}_\tau^{-1}[-d\Pi^{np}(s)/dQ^2]\}. \qquad (4)$$

In equation (4), $Q^2 \equiv -s$, and $\Pi^{np}(s)$ represents all correlation-function contributions from QCD-vacuum condensates as well as additional finite-correlation length contributions from the instanton background, i.e. "direct instanton contributions." The inverse Laplace transform in (4), corresponding to the Laplace-transform definition

$$\mathcal{L}_{Q^2}[f(\tau)] \equiv \int_0^\infty d\tau\, f(\tau)\, e^{-Q^2\tau}, \qquad (5)$$

is utilized to take advantage of the operator-product expansion of $\Pi^{np}$ in inverse powers of $Q^2$, and is easily understood via dispersion-relation methodology. Noting first that the singularities of $\Pi^h(s)$ [hadron poles and kinematic production-threshold branch cuts] must lie on the positive real s-axis, one finds that

$$(1/\tau)\, \mathcal{L}_\tau^{-1}[-d\Pi^{np}(s)/dQ^2]$$

$$= (1/\tau)\, \mathcal{L}_\tau^{-1}\Big[(1/\pi)\int_0^\infty ds\, \mathrm{Im}[\Pi^{np}(s)]/(s+Q^2)^2\Big] \qquad (6)$$

As is evident from (5), $\mathcal{L}_\tau^{-1}[1/(s+Q^2)^2] = \tau e^{-s\tau}$, which, upon substitution into (6) and (4), leads to a result consistent with duality between QCD $[\Pi^p(s) + \Pi^{np}(s)]$ and phenomenological hadronic physics $[\Pi^h(s)]$:

$$R_k^{QCD}(\tau) + \int_{s_0}^{\infty} ds\ (1/\pi)\ \mathrm{Im}[\Pi^p(s)]\ s^k\ e^{-s\tau}$$

$$= \int_0^{\infty} ds\ (1/\pi)\ \mathrm{Im}[\Pi^p(s) + \Pi^{np}(s)]\ s^k\ e^{-s\tau} \qquad (7)$$

Duality between $R_k^{QCD}(\tau)$ and $R_k^h(\tau)$ then follows via comparison of equations (7) and (2). We then find that the lowest lying resonance can be determined via the relationship:

$$\mathrm{Min}[R_1^{QCD}(\tau)/R_0^{QCD}(\tau)] \geq m_\ell^2 \qquad (8)$$

over an appropriate range of $\tau$ $[s_0^{1/2} > \tau^{-1/2} >> \Lambda_{QCD}]$.

*Laplace Sum-Rule Width Corrections to the Lowest-Lying Resonance Mass*

There is a need to go beyond the narrow resonance approximation if QCD sum rules exhibit sensitivity to resonance structures with non-zero widths. Such structures can not always be absorbed in the sum-rule continuum--even the lowest hadronic resonances may have substantial widths. For example, theoretical arguments exist [3,4] for the first pion-excitation to have a mass below 1 GeV, a floor for any reasonable estimate of the continuum threshold above which perturbative and hadronic QCD should coincide. Even if the first pion excitation state is identified with the Π(1300) resonance, whose mass pole is still likely to be below the continuum threshold, the width of this resonance may be as large as 600 MeV [5]. Moreover, the lowest isoscalar $0^+$ meson, if it exists at all [6], may have a width even larger than its mass [7], though arguments for a σ(550)-resonance with a somewhat more moderate width have also been recently advanced [8].

To gain qualitative insight into how nonzero resonance widths can effect QCD sum rule calculations, we can replace the δ-function of a resonance contribution to (1) with a rectangular pulse of unit area:

$$\delta(s - m^2) \;\rightarrow\; \wp_m(s,\Gamma) \;\equiv\; [\Theta(s-m^2+m\Gamma) - \Theta(s-m^2-m\Gamma)]/2m\Gamma. \tag{9}$$

Equation (9) defines a rectangular pulse centred at $s = m^2$ with full-width $\Delta s = 2m\Gamma$ and height $1/(2m\Gamma)$.

Let us consider how such an approximation to a lowest-lying resonance alters a QCD Laplace sum-rule determination of that resonance's mass. We assume in the spirit of the original formulation of QCD sum rules [1] that all but the lowest-lying ($\ell$) resonance is absorbed in the continuum. If we replace the delta function for the lowest-lying resonance with the pulse $\wp_m(s,\Gamma)$, we find from (2) that

$$R_0^h(\tau) \;=\; g_\ell \int_0^{s_0} ds\; \wp_m(s,\Gamma)\, e^{-s\tau} \;=\; g_\ell\, e^{-m^2\tau}\, \Delta_0(m,\Gamma,\tau), \tag{10}$$

$$R_1^h(\tau) \;=\; g_\ell \int_0^{s_0} ds\; \wp_m(s,\Gamma)\, s\, e^{-s\tau} \;=\; g_\ell m^2\, e^{-m^2\tau}\, \Delta_1(m,\Gamma,\tau), \tag{11}$$

with the functions $\Delta_{0,1}$ found from explicit evaluation of the integrals in (10) and (11):

$$\Delta_0(m,\Gamma,\tau) \;=\; \sinh(m\Gamma\tau)/(m\Gamma\tau) \tag{12}$$

$$\Delta_1(m,\Gamma,\tau) \;=\; \Delta_0(m,\Gamma,\tau)[1 + 1/(m^2\tau)] - \cosh(m\Gamma\tau)/(m^2\tau). \tag{13}$$

The results (10-13) assume that $\Gamma \leq m$, $s_0 \geq m^2 + m\Gamma$, so that the integration includes the entire resonance peak. Note also that

$$\lim_{\Gamma \to 0} \Delta_{0,1}(m,\Gamma,\tau) = 1,$$

consistent with the δ-function limit of a square pulse of infinitesimal width.

We see immediately from (10) and (11) that

$$m^2 = \frac{R_1^h(\tau)}{R_0^h(\tau)} \frac{\Delta_0(m,\Gamma,\tau)}{\Delta_1(m,\Gamma,\tau)}. \tag{14}$$

Since $R_1^h(\tau)/R_0^h(\tau)$ corresponds to $R_1^{QCD}/R_0^{QCD}$ by duality, and since this latter ratio corresponds to $m^2$ in the narrow resonance approximation ($\Gamma$=0), one can show from (12) and (13) that finite width effects will *increase* the masses of lowest-lying resonances extracted via Laplace sum rules:

$$\begin{aligned} m^2 &= [m^2]_{\Gamma=0} \cdot [\Delta_0(m,\Gamma,\tau)/\Delta_1(m,\Gamma,\tau)] \\ &= [m^2]_{\Gamma=0} \cdot [1 + \Gamma^2\tau/3 + O(\Gamma^4)]. \end{aligned} \tag{15}$$

This result should properly be regarded as a *lower-bound* on the magnitude of width contributions to $R_1/R_0$. The width $\Gamma$ appearing in (15) is, at present, defined via the rectangular pulse (9); it cannot be understood to correspond to a Breit-Wigner resonance width. Indeed, the narrow resonance approximation follows from the narrow-width limit of the Breit-Wigner resonance shape:

$$\begin{aligned} \mathrm{Im}[\Pi_\ell^h(s)] &= \mathrm{Im}[-g_\ell/(s-m^2+im\Gamma)] \\ &= g_\ell m\Gamma/[(s-m^2)^2 + m^2\Gamma^2] \underset{\Gamma\to 0}{\longrightarrow} \pi g_\ell \delta(s-m^2). \end{aligned} \tag{16}$$

If a resonance has a Breit-Wigner shape, then half of the total area of the resonance-peak

(considered as a function of s) is included in the range $m^2 - m\Gamma < s < m^2 + m\Gamma$. Since the unit-area rectangular pulse (9) has *all* of its area included in the range $m^2 - m\Gamma < s < m^2 + m\Gamma$, the result (15) is based upon a narrower pulse than the Breit-Wigner pulse of equivalent $\Gamma$, and is therefore likely to be an *underestimate* of the contribution of a Breit-Wigner resonance-width $\Gamma$ to the Laplace sum-rule determination of $m^2$.

For Laplace sum rules, a more quantitative estimate of resonance-width effects could be obtained by replacing the delta-functions in (1) with the Breit-Wigner peaks (16), and then substituting into the Laplace sum-rule definition (2). However, such an approach is subject to ambiguity. The Breit-Wigner shape has an infinite tail, and significant portions of that tail may extend above the continuum threshold $s_0$ or below the s=0 boundary into Euclidean momenta. Truncating such contributions would artificially exclude some of the integrated resonance peak. On the other hand, including such contributions leads to methodological contradictions with hadron physics. The Breit-Wigner shape, which itself stems from a linear approximation, can be modified for broad widths so as to vanish at s = 0 [9]. Even with such a modification, post-continuum contributions from the Breit-Wigner tail, whether included *or* truncated away, can be genuinely substantial for resonances with widths in excess of 100 MeV, and can be a source of *theoretical uncertainty* in Laplace sum-rule analyses of broad sub-continuum resonances.

Such uncertainty may be understood as a limitation on Laplace sum-rule methodology itself, particularly for channels in which more than one resonance lies below the continuum threshold. Non-lowest-lying resonances are expected to be less stable, and consequently, to be substantially broader than lowest-lying resonances. The I = 1 pseudoscalar channel has already been mentioned as an example of such a channel, and is discussed in the final two sections of this paper.[1]

---

[1] A QCD sum-rule treatment of the I = 0 scalar channel is even more likely to be problematical, in that this channel may be sensitive to not only the controversial sigma [$f_0$(400-1200)], but also as many as three other subcontinuum resonances [$f_0$(980), $f_0$(1370), and $f_0$(1500)], with $f_0$(1370) being a broad resonance.

In the section that follows, we first discuss how finite-energy sum rules can alleviate the resonance-width ambiguities described above. Unlike the case for Laplace sum rules, we demonstrate that the contribution of a non-narrow resonance to the first two finite-energy sum rules ($F_0$ and $F_1$) is *independent* of that resonance's width, provided the resonance peak is both symmetric and entirely below the continuum threshold.

These same two sum rules are also less sensitive to higher dimensional condensates than corresponding Laplace sum rules, thereby lowering the number of nonperturbative QCD-vacuum parameters required to enter a sum-rule analysis. In Section III, we demonstrate that $F_0$ and $F_1$ are essentially decoupled (to leading order in $\alpha_s$) from dimension > 6 QCD-vacuum condensates containing one or more $\overline{q}q$ pairs. For the specific channel pertinent to pseudoscalar resonances, this suppression is shown to occur even for dimension > 4; the dimension-6 condensate $\langle\alpha_s(\overline{q}q)^2\rangle$ is seen (by virtue of a third order pole at $Q^2 = 0$) not to enter $F_{0,1}$, even though it *is* known to enter the corresponding Laplace sum rules $R_{0,1}$. Insensitivity of the first two finite-energy sum rules to multiple gluon condensates [condensates which do not have any $\overline{q}q$ pairs] of dimension $\geq 6$ is discussed at length in Section IV, including the operator-mixing that serves to suppress the dimension-6 gluon condensate.

We then focus on QCD sum rules for the I = 1 pseudoscalar mesons, as noted earlier, as an example of a "problem" channel with broad subcontinuum resonance contributions. This channel has long been understood to be subject to instanton contributions. In Section V, we extract the direct single-instanton contribution to the finite-energy sum rule $F_1$ for this channel. In Section VI, we relate $F_1$ to the phenomenology of this channel, which has been used by others [10,11,12] to obtain a fairly large lower bound on the light quark mass ($m_u + m_d \geq 20 \pm 5$ MeV). We demonstrate how direct single-instanton contributions to $F_1$ not only lower the bound for the estimated quark mass, but also lower the estimated contribution of higher-mass resonances to the sum rule to values of $r_i$ [$\equiv (F_i M_i^2/f_\pi m_\pi^2)^2$] more consistent with present phenomenological expectations.

Our manuscript also possesses four detailed appendices pertinent to a full methodological understanding of finite energy sum rules. In Appendix A an exact expression for the imaginary part of a function that arises in closed fermion loop contributions to correlation functions [i.e. in purely perturbative and multiple-gluon condensate contributions] is extracted. In Appendices B and C, the two-gluon condensate contributions to $F_0$ and $F_1$ are respectively calculated from that condensate's *exact* (as opposed to leading-order in $m_q$) one-loop contribution to the longitudinal component of the axial vector correlation function. This contribution is shown to arise entirely from a net branch singularity for $s \geq 4m^2$. The absence of net pole contributions at s = 0, as well as the cancellation of infrared singularities arising from integration of the exact expression along the branch cut against those arising from integration around the branch-cut terminus at $s = 4m^2$, are also demonstrated explicitly. All of these results (including the singularity structure described above) are applicable to the two-gluon condensate contributions to the finite energy sum rules associated with the scalar, vector and the transverse component of the axial vector correlation functions. These sum rule contributions are itemized in Appendix D. The explicit cancellation of quark-mass singularities via operator mixing is also demonstrated for channels in which they naively occur.

## II. Nonzero Resonance Widths and Finite-Energy Sum-Rules

*Finite Energy Sum Rules and Higher-Mass Resonances*

For a given current-correlation function $\Pi(s)$, the finite-energy sum-rules (FESR's) $F_k(s_0)$ are defined here to be the integrals [13]

$$F_k(s_0) \equiv (1/2\pi i) \int_{C(s_0)} ds\, s^k\, \Pi(s), \tag{17}$$

where the contour $C(s_0)$ is an open circle of radius $s_0$ in the complex s-plane that does not cross the real s-axis [Fig. 1a]. The parameter $s_0$ is understood to be the continuum threshold discussed in the previous section. For the hadronic contribution to the FESR's $F_k^h(s_0)$, the contour $C(s_0)$ can be distorted into a line running below and above the physical singularities on the positive real s-axis [Fig. 1b]:

$$F_k^h(s_0) = (1/\pi) \int_0^{s_0} ds\, s^k\, \mathrm{Im}[\Pi^h(s)]. \qquad (18)$$

In the narrow resonance approximation (1), one finds that

$$F_k^h(s_0) \;=\; \sum_r g_r\, m_r^{2k} \;\equiv\; \sum_r [F_k^h(s_0)]_r \qquad (19)$$

an expression that differs from (3) only in that higher-mass sub-continuum resonances are no longer exponentially suppressed. This is a positive feature of the FESR approach, if one is seeking to use sum rules to obtain information about such resonances.

*Insensitivity of $F_{0,1}$ to Symmetric-Peak Resonance Widths*

To examine finite width effects, let us first replace the delta-functions of (1) with the finite-width rectangular pulses (9). As long as $s_0 \geq m_r^2 + 2m_r\Gamma_r$, the contribution of such a pulse to $F_0$ is clearly the same as that of a delta-function, since $F_0$ is sensitive only to peak-area:

$$[F_0^h(s_0)]_r \quad \to \quad \int_0^{s_0} ds\, g_r \wp_{m_r}(s,\Gamma_r) = g_r \qquad (20)$$

Remarkably, the $F_1$ sum-rule is *also* insensitive to the width of the rectangular pulse:

$$[F_1^h(s_0)]_r \quad \to \quad \int_0^{s_0} ds\, s\, g_r \wp_{m_r}(s,\Gamma_r) = [g_r/(2m_r\Gamma_r)] \int_{m_r^2-m_r\Gamma_r}^{m_r^2+m_r\Gamma_r} ds\, s \quad = \quad g_r m_r^2 \qquad (21)$$

The final result of (21) is identical to the contribution to $[F_1^h(s_0)]_r$ obtained from the narrow resonance approximation, with $\wp_{m_r}(s,\Gamma_r)$ replaced by $\delta(s-m_r^2)$. The results (20) and (21) are to be contrasted with the width-dependence exhibited in (10) and (11) for corresponding Laplace sum rules.

Moreover, the width-independence of the first two FESR's obtained above is *not* an artifact of the rectangular pulse approximation for non-zero width resonances. Any symmetric resonance peak $\mathbb{R}_m(s)$ centred at $m^2$ can be represented as a sum over variable-width unit-area rectangular pulses $\wp_m(s,\Gamma')$ centred at $s = m^2$:

$$\mathbb{R}_m(s) \quad = \quad \int_0^{\Gamma_{max}} d\Gamma' \;\; f(\Gamma') \; \wp_m(s,\Gamma') \qquad (22)$$

In (22), $f(\Gamma')$ is just the weighting assigned to the unit-area rectangular pulse with width $\Gamma'$. For example, the Breit-Wigner shape (16) can be expressed in the form of (22) by converting a Riemann sum of infinitesimally thin pulses into an integral [Fig. 2]:

$$m\Gamma/[(s-m^2)^2+m^2\Gamma^2]$$

$$= \quad \lim_{n\to\infty} \quad \{(2/n) \; \sum_{j=1}^{n} \; \sqrt{(n/j)-1} \; \wp_m(s, \sqrt{(n/j)-1}\;\Gamma)\}$$

$$= \; 2 \; \int_0^1 dy \;\; \sqrt{(1-y)/y} \;\; \wp_m(s, \sqrt{(1-y)/y}\;\Gamma)$$

$$= \quad 4\Gamma \int_0^\infty \frac{d\Gamma' \, (\Gamma')^2}{(\Gamma^2+\Gamma'^2)^2} \;\; \wp_m(s,\Gamma') \qquad (23)$$

Assuming the peak $\mathbb{R}_m(s)$ has an area normalized to $\pi$, consistent with $\mathbb{R}_m(s) \to \pi\,\delta(s-m^2)$ in the narrow resonance limit [e.g. eq. (16)], one finds that

$$\pi = \quad \int_0^{s_0} ds \; \mathbb{R}_m(s) = \int_0^{\Gamma_{max}} d\Gamma' \;\; f(\Gamma') \qquad (24)$$

provided $s_0 > m^2 + m\Gamma_{max}$.[2] Consequently, one can use (21) and (24) to demonstrate that replacing factors of $\pi\ \delta(s-m^2)$ in (1) with $\mathbb{R}_m(s)$ will not alter narrow-resonance approximation predictions (19) for $F_0$ and $F_1$:

$$[F_0^h(s_0)]_r \quad \rightarrow \quad \int_0^{s_0} ds\ (1/\pi)\ g_r\ [\mathbb{R}_m(s)]_r = g_r \tag{25}$$

$$[F_1^h(s_0)]_r \quad \rightarrow \quad \int_0^{s_0} ds\ s\ (1/\pi) g_r [\mathbb{R}_m(s)]_r$$

$$= \quad (1/\pi) \int_0^{\Gamma_{max}} d\Gamma'\ \ f(\Gamma') \int_0^{s_0} ds\ s\ g_r\ \wp_{m_r}(s,\Gamma') = g_r m_r^2. \tag{26}$$

Thus we see that the finite-energy sum-rules $F_0$ and $F_1$ are impervious to resonance-width effects, provided the resonance in question is a symmetric peak that is entirely below the continuum threshold $s_0$. Consequently, we observe that these sum rules are particularly well-suited for an analysis of non-lowest-lying subcontinuum resonances. As remarked earlier, the contributions (19) of such resonances to $F_0$ and $F_1$ are not exponentially suppressed, as is the case with Laplace sum-rules (3). Since such resonances are unstable, and therefore broad, it is phenomenologically useful that their contributions to $F_0$ and $F_1$ are unaffected by their decay widths. It should be noted this property does not apply to higher finite energy sum rules; sum rules $F_k$ *do* exhibit width dependence if $k \geq 2$. Consequently, we will restrict our discussion henceforth to the properties of $F_{0,1}$ sum rules.

---

[2] For the theoretical Breit-Wigner shape, problems associated with the infinite tail of the resonance extending past the continuum threshold $s_0$ have already been mentioned. One approximation which truncates the Breit-Wigner shape symmetrically is to choose $\Gamma_{max} = 2m$. This approximation makes sense, however, only if $\Gamma << 2m \leq s_0$.

## III. FESR Suppression of Higher-Dimensional Condensates with One or More $\overline{q}q$-Pairs

*The General Case*

The operator-product expansion (OPE) for a dimension-2 two-current correlation function $\Pi(s)$ can be expressed at Euclidean momenta $Q^2 \equiv -s > 0$ in terms of QCD-vacuum condensates as follows:

$$\Pi(-Q^2) = C_p(Q^2) + C_{\overline{q}q}(Q^2)\langle m_q\overline{q}q\rangle + C_{G^2}(Q^2)\langle\alpha_s G^2\rangle + C_M(Q^2)\langle\overline{q}G\cdot\sigma q\rangle$$
$$+ C_{G^3}(Q^2)\langle\alpha_s G^3\rangle + C_{(\overline{q}q)^2}(Q^2)\langle\alpha_s(\overline{q}q)^2\rangle + ... \qquad (27)$$

To leading order in $\alpha_s$, the OPE coefficients $C_n(Q^2)$ of an n-dimensional condensate $\langle O_n\rangle$ are of the general form

$$C_n(Q^2) \quad = \quad \sum_j [A_j + B_j \ln(Q^2/\mu^2)]\, m_q^j/Q^{j+n-2} \qquad (28)$$

To avoid mass singularities, the index j is restricted to zero and even positive integers if n is even, and to odd positive integers if n is odd. To leading order in $\alpha_s$, contributions to (27) from condensates containing at least one fermion-antifermion pair necessarily correspond to diagrams with broken loops [Fig. 3], and for such diagrams $B_p = 0$; logarithms from integrations over closed-loop momenta do not occur. For example, to leading order in $\alpha_s$, the $\langle m_q\overline{q}q\rangle$ contribution [Fig. 3a] to the longitudinal component $\Pi^L(s)$ of the axial-vector current correlation function,

$$(g_{\mu\nu} - p_\mu p_\nu/p^2)\Pi^T(p^2) + (p_\mu p_\nu/p^2)\Pi^L(p^2) \equiv i\int d^4x\, e^{ip\cdot x}\langle 0|T\, j_{\mu 5}(x)\, j_{\nu 5}(0)|0\rangle, \qquad (29)$$

$[j_{\mu 5} \equiv \overline{u}\gamma_\mu\gamma_5 d]$ is given by [14]

$$C^L_{\overline{q}q}(Q^2) = (2/m_q^2)[1 - (1 + 4m_q^2/Q^2)^{1/2}] = -4/Q^2 + 4m_q^2/Q^4 - 8m_q^4/Q^6 + ... \qquad (30)$$

If in (28) $B_j = 0$ for all j, the definition (17) implies that the FESR's $F_0$ and $F_1$ are (respectively) sensitive *only* to first and second order poles at $Q^2 = 0$:

$$[F_0^L(s_0)]_{\bar{q}q} = (1/2\pi i)\int_{C(s_0)} ds\, C_{\bar{q}q}^L(-s)\langle m_q\bar{q}q\rangle = -4\,\langle m_q\bar{q}q\rangle \tag{31}$$

$$[F_1^L(s_0)]_{\bar{q}q} = (1/2\pi i)\int_{C(s_0)} ds\, s\, C_{\bar{q}q}^L(-s)\langle m_q\bar{q}q\rangle = -4m_q^2\,\langle m_q\bar{q}q\rangle \tag{32}$$

Thus, if $C_n(Q^2)$, the OPE coefficient of a condensate $\langle O_n\rangle$, is restricted to inverse powers of $Q^2$, then n must be less than or equal to 6 for that condensate to contribute to $F_0$ or $F_1$. If $n > 6$, then $n+j-2 \geq 6$ and the leading OPE contribution to (28) is at least a third order pole at $Q^2 = 0$, which cannot contribute to $F_0$ or $F_1$.

*Higher-Dimensional Fermionic Condensates in the Pseudoscalar Channel*

For the particular case of the longitudinal component (L) of the axial-vector correlation function, which is coupled to pion-resonance states, there is an additional chiral symmetry constraint that $C_n^L(Q^2) \to 0$ as $m_q \to 0$, in which case $j \geq 1$ for all coefficients of condensates that fail to vanish in the chiral limit. As a consequence, one can show to leading order in $\alpha_s$ that the n = 6 condensate $\langle\alpha_s(\bar{q}q)^2\rangle$ cannot contribute to $F_0$ or $F_1$, as its leading contribution is necessarily a third-order pole at $Q^2 = 0$ [1,10]:

$$C_{(\bar{q}q)^2}^L(Q^2) = -448\pi\, m_q^2\, \alpha_s/27Q^6 + O(m_q^4/Q^8). \tag{33}$$

Similarly, $F_0$ and $F_1$ are found to be insensitive to the (n = 5) mixed condensate $\langle\bar{q}G\cdot\sigma q\rangle$. The relevant contribution to the longitudinal component of the axial-vector correlator is also seen to involve only third-and-higher order poles at $Q^2 = 0$ [15]:

$$C_M^L(Q^2) = -(1-v)^3/2m_q^3 v = 4m_q^3/Q^6 - 20m_q^5/Q^8 + \ldots, \tag{34}$$

$$v \equiv (1 + 4m_q^2/Q^2)^{1/2}. \tag{35}$$

Thus the leading contributions to $F_0$ and $F_1$ sum rules in this channel do not involve any condensates with quark-antiquark pairs except $\langle m_q \bar{q} q\rangle$. The $F_0$ and $F_1$ sum rules in other channels can also involve the n=5 mixed condensate $\langle \bar{q} G \cdot \sigma q\rangle$ and the n=6 condensate $\langle \alpha_s (\bar{q} q)^2 \rangle$ [we are assuming vacuum-saturation], but no other condensates containing quark-antiquark pairs, as all other such condensates are of dimension greater than 6.

## IV. Purely Gluonic Contributions to $F_0$ and $F_1$

*Purely Perturbative Gluon-Loop Contributions*

For two-current correlation functions, the suppression of leading-order contributions from n > 6 condensates applies only to those operators whose leading contribution in $\alpha_s$ does not involve a closed perturbative loop. However, all condensates involving gluons necessarily are generated from the closed-loop vacuum polarization diagram [Fig. 4], and such diagrams are characterized by nonzero coefficients $B_j$ in the OPE expansion (28). The contribution of such logarithmic terms in (28) to the FESRs $F_0$ and $F_1$ can be obtained from the general relation [$Q^2 \equiv -s$; D is an integer]

$$\int_{C(s_0)} ds \ln(Q^2)/(Q^2)^D = -2i\pi(-1)^D s_0^{1-D}/(1-D); \quad D \geq 2 \tag{36}$$

However, a more precise evaluation of the contributions of closed-loop OPE coefficients necessarily involves the one-loop momentum integral[3]

$$X(v) \equiv (1/v^2) \int_0^1 dx \ln[1 - s\, x(1-x)/m_q^2 - i|\varepsilon|] + 2/v^2, \tag{37}$$

$$v \equiv (1 - 4m_q^2/s)^{½}. \tag{38}$$

[3] In eqs. (37-9) we are utilizing the notation of ref. [15].

For Euclidean momenta ($s < 0$), one finds $X(v)$ to be the real function

$$X(v) = (1/v)\ \ln[(1+v)/(v-1)]. \tag{39}$$

For Minkowskian momenta ($s > 0$), $X(v)$ develops an imaginary part above the quark-antiquark kinematic production threshold, as discussed in Appendix A:

$$X(v) = (1/v)\ \{\ln[(1+v)/(1-v)] - i\pi\},\ s > 4m_q^2. \tag{40}$$

The result (40) facilitates the sum-rule determination of closed loop contributions to $F_{0,1}$. For example, the one-loop purely perturbative contribution [Fig. 4a] to the longitudinal component of the axial-vector current correlator (29) is given by [15]

$$C_p^L[v] = (-3m_q^2/2\pi^2)[v^2 X(v) + \text{divergent constant}]. \tag{41}$$

The contribution of (41) to $F_{0,1}$ is easily obtained via a distortion of the contour $C(s_0)$ to that in Fig. 1b:

$$[F_0^L(s_0)]_p = (1/\pi) \int_0^{s_0} ds\ \mathrm{Im}\{C_p^L[v]\}$$

$$= (3m_q^2/2\pi^2) \int_{4m_q^2}^{s_0} ds\ (1 - 4m_q^2/s)^{1/2}$$

$$= (3m_q^2/2\pi^2)\{s_0(1-4m_q^2/s_0)^{1/2} + 2m^2\ \ln|[1-(1-4m_q^2/s_0)^{1/2}]/[1+(1-4m_q^2/s_0)^{1/2}]|\} \tag{42}$$

$$[F_1^L(s_0)]_p = (1/\pi) \int_0^{s_0} ds\ s\ \mathrm{Im}\{C_p^L[v]\}$$

$$= (3m_q^2/2\pi^2) \int_{4m_q^2}^{s_0} ds\ s\ (1 - 4m_q^2/s)^{1/2}$$

$$= (3m_q^2/4\pi^2)(s_0^2 - 2m^2 s_0)(1 - 4m_q^2/s_0)^{1/2}$$

$$+ (3m_q^6/\pi^2) \ln|[1-(1-4m_q^2/s_0)^{1/2}]/[1+(1-4m_q^2/s_0)^{1/2}]|\}. \qquad (43)$$

We note that the results (42) and (43) are *exact* expressions obtained from the one-loop expression (41). To leading order in the quark mass $m_q$, one finds from (42) and (43) that

$$[F_0^L(s_0)]_p = (3m_q^2 s_0/2\pi^2) + O(m_q^4), \qquad (44)$$

$$[F_1^L(s_0)]_p = (3m_q^2 s_0^2/4\pi^2) + O(m_q^4). \qquad (45)$$

*Two-Gluon Condensate Contributions to $F_{0,1}$*

The OPE coefficient $C_{G^2}(Q^2)$ is extracted from the OPE coefficient $E_{G^2}(Q^2)$ in the "normal-ordered basis" [i.e., the "heavy quark" coefficients listed in Appendix B of ref. 15] as follows :

$$C_{G^2}(Q^2) = E_{G^2}(Q^2) + (1/12\pi)C_{\bar{q}q}(Q^2) - (m_q/2\pi)\ln(m_q^2/\mu^2)\ C_M(Q^2). \qquad (46)$$

The linear combination (46) represents the coefficient in the "minimally-subtracted basis," which is chosen so as to avoid mass-singularities [16]. For example, in the normal-ordered basis the coefficient of $\langle\alpha_s G^2\rangle$ for the longitudinal component of the axial-vector correlator is [15]

$$E^L_{G^2}(Q^2) = (-1/96\pi Q^2)[16m_q^4(3+9v^2)X(v)/(v^4Q^4)] + [1/(48\pi v^4Q^2)][9v^4+ 4v^2+ 3] \tag{47}$$

which generates an expansion

$$E^L_{G^2}(Q^2) = 1/3\pi Q^2 - 5m_q^2/6\pi Q^4 + (m_q^4/\pi Q^6)[13/3 + 2\ln(m_q^2/Q^2)] + O(m_q^6/Q^8). \tag{48}$$

The leading term on the right hand side of (48) does not vanish as $m_q \to 0$, despite the chiral invariance of $\langle\alpha_s G^2\rangle$. Moreover, the right hand side has the quark mass appear in the logarithm, which could (in principle) lead to a large logarithm after subtractions. The change of basis (46) eliminates both problems, as is evident from direct substitution of (48), (34) and (30) into (46):

$$C_{G^2}(Q^2) = - m_q^2/2\pi Q^4 + (m_q^4/\pi Q^6)[11/3 - 2\ln(Q^2/\mu^2)] + O(m_q^6/Q^8). \tag{49}$$

The result (49) is consistent with the general form (28), although the recipe (46) requires further modification if $O[m_q^6\ln(m_q^2/Q^2)/Q^8]$ terms are to be eliminated. It is worth noting that the change of basis (46) differs from an operator redefinition proposed on chiral symmetry grounds in ref. [15] only by the presence of the final $C_M(Q^2)$ term, which has already been shown not to affect the contour integrals leading to $F_0$ and $F_1$. In Appendices B and C, the full contribution of $C^L_{G^2}(Q^2)$ to the $F_{0,1}$ sum rules for the longitudinal component of the axial-vector correlator is determined to all orders in $m_q$ by careful consideration of the $C(s_0)$ contour. However, contributions to $F_0$ and $F_1$ from the $\langle\alpha_s G^2\rangle$ condensate can be evaluated to $O(m_q^4)$ from application of (36) and the Cauchy residue theorem to (49):

$$[F^L_0(s_0)]_{\langle\alpha_s G^2\rangle} = (1/2\pi i)\langle\alpha_s G^2\rangle\int_{C(s_0)} ds\, C^L_{G^2}(-s) = (m_q^4/\pi s_0^2)\langle\alpha_s G^2\rangle \tag{50}$$

$$[F^L_1(s_0)]_{\langle\alpha_s G^2\rangle} = (1/2\pi i)\langle\alpha_s G^2\rangle\int_{C(s_0)} ds\, s\, C^L_{G^2}(-s) = [m_q^2/2\pi + 2m_q^4/\pi s_0]\langle\alpha_s G^2\rangle. \tag{51}$$

As in (46), the OPE coefficient $C_{G^3}(Q^2)$ in the minimally-subtracted basis can be extracted from the coefficient $E_{G^3}(Q^2)$ in the normal-ordered "heavy quark" basis [15,17]:

$$C_{G^3}(Q^2) = E_{G^3}(Q^2) + [1/(360\pi m_q^2)]\, C_{\overline{q}q}(Q^2) + [1/(12\pi m_q)]\, C_M(Q^2). \qquad (52)$$

This change of basis once again eliminates leading-order mass-singularities. To see this, we demonstrate application of (52) to FESR's by once again considering the relevant contributions to the longitudinal component of the axial-vector current correlation function. The OPE coefficient $E^L_{G^3}(Q^2)$ is given by [15]

$$E^L_{G^3}(Q^2) = \frac{-m_q^4}{24\pi Q^8 v^8}\, X(v)[7 + 23v^2 + 13v^4 + 5v^6] + \frac{1}{2880\pi Q^4 v^8 (1-v^2)}\,[105 + 65v^2 - 494v^4 + 266v^6 + 5v^8 - 75v^{10}], \qquad (53)$$

which generates the following expansion in inverse powers of $Q^2$:

$$E^L_{G^3}(Q^2) = \frac{1}{\pi Q^2}\left[\frac{1}{90 m_q^2} - \frac{1}{90 Q^2} - \frac{14 m_q^2}{45 Q^4} + O(m_q^4)\right] \qquad (54)$$

The leading term of (54) diverges as $m_q \rightarrow 0$, an explicit mass singularity. The next-to-leading term fails to vanish in the chiral limit. However, both of these terms, as well as the explicit $O(m_q^2)$ term in (54) cancel in (52) against corresponding terms from $C^L_{\overline{q}q}$ (30) and $C^L_M$ (34). Consequently, the OPE coefficient $C^L_{G^3}$ is explicitly $O(m_q^4)$, and is therefore suppressed relative to $C^L_{G^2}$.

The suppression of $C_{G^3}$ relative to $C_{G^2}$ in the operator product expansion appears to be a general property [17,18]. Corresponding factors of $C_{G^2}$ and $C_{G^3}$ for the scalar and vector current correlation functions, as well as for the transverse component of the axial-vector correlation function, as extracted via (46) and (52) from the "heavy quark" expressions in [15], are also seen to exhibit suppression by $m_q^2$:

By contrast, the dimension-8 contributions to scalar, pseudoscalar and vector correlation functions (which in our conventions are defined to have dimensions of mass squared) are shown in ref.[19] to be of the form $[A_0 + B_0 \ln(Q^2/\mu^2)]\langle G^4\rangle/Q^6$, where $A_0$ and $B_0$ are numerical: suppression by $m_q^2$ does not seem to occur. For the *longitudinal* component of the axial-vector correlator, which picks up a factor of $m_q^2/Q^2$ relative to the pseudoscalar correlator, a dimension-8 contribution to $F_1^L$ will then be proportional [via (36)] to $m_q^2 B_0\langle G^4\rangle/s_0^2$. Such a contribution will be small compared to that of the dimension-4 condensate $\langle G^2\rangle$ [eq. (51)] provided $B_0\langle G^4\rangle$ is small compared to $\langle G^2\rangle s_0^2$, suggesting, in the absence of $m_q^2$-suppression factors, that any further suppression of 2d-dimensional gluon condensates to FESR's is contingent upon the ratio $\langle G^{2d}\rangle/s_0^d$ being small. Such a small ratio can be anticipated via dimensional and factorization arguments [e.g. $\langle G^2\rangle < s_0$].

## V. Direct Single-Instanton Contributions to $F_1^L$

In the instanton liquid model, the direct single-instanton contribution to the $R_0$ Laplace sum rule (2) for the pseudoscalar (P) correlation function has been found to be [20]

$$R_0^P(\tau) \equiv (1/\pi)\int_0^\infty \mathrm{Im}\{[\Pi^P(s)]_{inst}\}\, e^{-s\tau}\, ds$$

$$= \frac{3\rho^2}{8\pi^2\tau^3}\, e^{-\rho^2/2\tau}\, [K_0(\rho^2/2\tau) + K_1(\rho^2/2\tau)], \quad (55)$$

where $\rho$ [= 1/(600 MeV)] is the instanton-size parameter. Since the pseudoscalar correlator is related to the longitudinal component (L) of the axial-vector correlator by

$$\Pi^L(s) = 4m_q^2\Pi^P(s)/s, \tag{56}$$

we see that the instanton contribution $\mathscr{F}_1$ to the corresponding FESR $F_1^L$ is[4]

$$\mathscr{F}_1(s_0) = (1/\pi)\int_0^{s_0} \mathrm{Im}\{[\Pi^L(s)]_{inst}\}\ s\ ds$$

$$= (4m_q^2/\pi)\int_0^{s_0} \mathrm{Im}\{[\Pi^P(s)]_{inst}\}\ ds, \tag{57}$$

which is related via Laplace transformation to the function (55) for $R_0^P(s)$ as follows:

$$\mathscr{F}_1'(t) = (4m_q^2/\pi)\ \mathrm{Im}\{[\Pi^P(t)]_{inst}\}, \tag{58}$$

$$\mathfrak{L}[\mathscr{F}_1'(t)] \equiv \int_0^\infty \mathscr{F}_1'(t)\ e^{-st}\ dt = 4m_q^2\ R_0^P(s) = s\ \mathfrak{L}[\mathscr{F}_1(t)] - \mathscr{F}_1(0). \tag{59}$$

We see from (57) that $\mathscr{F}_1(0) = 0$, and find that

$$\mathscr{F}_1(t) = \mathfrak{L}^{-1}[4m_q^2\ R_0^P(s)/s]$$

$$= \mathfrak{L}^{-1}\left[\frac{3\rho^2 m_q^2}{2\pi^2 s^4} e^{-\rho^2/2s}\ [K_0(\rho^2/2s) + K_1(\rho^2/2s)]\right]. \tag{60}$$

---

[4] From (56), the direct single-instanton contributions to $F_0^L$ and $F_1^L$ are both $O(m_q^2)$. This implies that the instanton contribution to $F_0^L$ is small in comparison to (31), the leading (quark-condensate) contribution to $F_0^L$, which is why we are only concerned here with instanton contributions to $F_1^L$.

Our use of the variables s and t is to retain consistency with standard Laplace transform conventions; the variable t will ultimately be identified with the continuum threshold $s_0$, and the variable s corresponds to the Borel parameter $\tau$ in (55) [as defined in (2)]. The inverse transform of (60) may be obtained from the asymptotic expansions of $K_0$ and $K_1$ [21]:

$$K_0(z) + K_1(z) = (\pi/2z)^{1/2} e^{-z} \left[2 + 1/(4z) - 3/(64z^2) + 15/(512z^3) \ldots\right] \tag{61}$$

$$\mathcal{F}_1(t) = \frac{3m_q^2\rho}{\pi^{3/2}}\left[\mathcal{L}^{-1}\left(s^{-7/2}e^{-\rho^2/s}\right) + \frac{1}{4\rho^2}\mathcal{L}^{-1}\left(s^{-5/2}e^{-\rho^2/s}\right) - \frac{3}{32\rho^4}\mathcal{L}^{-1}\left(s^{-3/2}e^{-\rho^2/s}\right) + \frac{15}{128\rho^6}\mathcal{L}^{-1}\left(s^{-1/2}e^{-\rho^2/s}\right) + \ldots\right] \tag{62}$$

Using (62) and replacing t with $s_0$, we find that

$$\mathcal{F}_1(s_0) = \frac{3m_q^2}{\pi^2\rho^4}\ G(2\rho s_0^{1/2}), \tag{63}$$

where the function $G(2\rho s_0^{1/2})$ is defined via

$$G(w) \equiv \{[-w^2/4 + 25/32 + O(1/w^2)]\sin(w) + [-7w/8 + 15/(64w) + O(1/w^3)]\cos(w)\}. \tag{64}$$

The results (62-64) are not useful unless $w \to 2\rho s_0^{1/2} > 1$. Since $s_0^{1/2}$ is generally expected to be at least 1 GeV, the expansion in large w is appropriate and useful [$\rho^{-1} \cong 0.6$ GeV]. In the large $s_0$ limit, the leading perturbative contribution to $F_1^L$ [eq.(45)] dominates the instanton contribution, which is at most linear in $s_0$ (63-4). However, for values of $s_0$ near 1 GeV$^2$, the instanton contribution is shown in the next section to be larger than the perturbative contribution, with phenomenological implications for the light quark mass.

## VI. Discussion: FESR's in the Pseudoscalar Channel and the Light-Quark Mass

An old [22] and ongoing [23] controversy in sum rule applications concerns the failure of the field-theoretical content of the QCD sum rules to saturate the pseudoscalar channel. The essence of this problem is evident from a qualitative examination of the $R_0$ and $R_1$ Laplace sum rules for the longitudinal component of the axial-vector current correlation function, as defined in (2) and (4). For suitable values of the Borel parameter $\tau$ ($M \equiv \tau^{-1/2} >> m_\pi$), one finds

that

$$R_0 = \quad f_\pi^2 m_\pi^2 + \sum_{M_i^2 < s_0} F_i^2 M_i^2 \exp(-M_i^2\tau) \; = \; -4\langle m_q \overline{q} q\rangle + O(m_q^2), \tag{65}$$

a result consistent with the current-algebra GMOR relationship $f_\pi^2 m_\pi^2 = -4\langle m_q \overline{q} q\rangle$ [24] as long as the subsequent subcontinuum resonances in the summation on the hadronic side of (65) are either sufficiently heavy ($M_i^2 >> 1/\tau >> m_\pi^2$), or their decay constants $F_i^2$ are sufficiently small ($F_i^2 << f_\pi^2 m_\pi^2/M_i^2$). The leading field-theoretical contribution to the $R_1$ sum rule, however, is quadratic in the quark mass [1,10]:

$$R_1 = \quad f_\pi^2 m_\pi^4 + \sum_{M_i^2 < s_0} F_i^2 M_i^4 \exp(-M_i^2\tau) \; = \; m_q^2[-4\langle m_q \overline{q} q\rangle + 3/(2\pi^2\tau^2) + \langle \alpha_s G^2\rangle/2\pi$$
$$+ 448\pi\tau\langle \alpha_s(\overline{q}q)^2\rangle/27 + ...]. \tag{66}$$

Naively, the field-theoretical content of (66) is of order $m_q^2$ times the field-theoretical content of (65), whereas the hadronic content of (66) is at least of order $m_\pi^2$ times the hadronic content of (65), suggesting that $m_q$ and $m_\pi$ are comparable. A thorough treatment of QCD contributions to (66) still yields substantially larger values of the light quark mass [10-12] than are anticipated from other phenomenology [5], as already noted in the Introduction to this paper.

This mismatch in scale [i.e., $R_1^h/R_0^h \sim m_\pi^2$; $R_1^{QCD}/R_0^{QCD} \sim m_q^2$] superficially characterizes the FESR's $F_0^L$ and $F_1^L$ as well. However, these FESR's provide a much cleaner framework for extracting limits on $m_q$, enabling one to avoid the large-width modifications to the hadronic-resonance content of (66), as discussed in Sections II and III, as well as higher-dimensional condensate contributions (including that of $\langle \alpha_s(\overline{q}q)^2\rangle$) to the field-theoretical content of (66), as is also discussed in Sections IV and V.

Direct single-instanton contributions (55) to the Laplace sum rule have been argued in a number of places [1,20,23] to be necessary for the saturation of the pseudoscalar channel. If we incorporate such contributions (63,64) into the FESR $F_1^L$, in conjunction with the (width-independent) hadronic contributions (19) as well as the leading [$O(m_q^2)$] field theoretical contributions (32,45,51), we find that

$$F_1^L = f_\pi^2 m_\pi^4 + \sum_{M_i^2 < s_0} F_i^2 M_i^4 = m_q^2[-4\langle m_q \bar{q} q\rangle + \langle \alpha_s G^2\rangle/2\pi + 3s_0^2/4\pi^2$$
$$+ (3/\pi^2\rho^4)\, G(2\rho s_0^{1/2}) + O(m_q)]. \tag{67}$$

For each subcontinuum pion-excitation state, we define the parameter

$$r_i \equiv (F_i^2 M_i^4)/(f_\pi^2 m_\pi^4). \tag{68}$$

We can then rearrange (67) to obtain the following relationship for the light-quark mass:

$$m_q^2 = \frac{f_\pi^2 m_\pi^4 (1 + \sum r_i)}{A + [G(w) + w^4/64]B}, \tag{69}$$

where...

1) ... the summation is understood to be over only those resonance peaks below the continuum threshold ($M_i^2 < s_0$);
2) ... the dependence on the continuum-threshold $s_0$ enters through the variable

$$w \equiv 2\rho s_0^{1/2}; \tag{70}$$

3) ... the function G(w) is given by (64); and
4) ... the constants A and B are given by

$$A \equiv -4\langle m_q \bar{q} q\rangle + \langle \alpha_s G^2\rangle/2\pi, \tag{71}$$

$$B \equiv 3/(\pi^2\rho^4). \tag{72}$$

The relationship (69) should retain approximate validity as $s_0$ increases to include (*in full*, as noted in Section III) the resonance peaks of additional pion-excitation states. In particular, one would expect the contribution from $\Pi(1300)$, the first pion-excitation ($M = 1300 \pm 100$ MeV, $\Gamma = 200 - 600$ MeV [5]) to be fully subcontinuum if $s_0 > 4\ GeV^2$. Possible additional contributions may accrue in full from $\Pi(1770)$ and $X(1830)$ at even larger values of $s_0$.

Using standard parameter values [$\langle m_q \bar{q} q\rangle = -f_\pi^2 m_\pi^2/4$, $\langle \alpha_s G^2\rangle = 0.045\ GeV^4$, $\rho^{-1} = 600$ MeV], one can then estimate the following numerical lower bound on the quark mass from (69):

$$m_q = \mu(w)\sqrt{1 + \Sigma\, r_i} > \mu(w)\sqrt{1 + r_1 \Theta(s_0 - 4\ GeV^2)}, \tag{73}$$

$$\mu(w) = \frac{2.6\ \text{MeV}}{\{0.0075 + 0.039[G(w) + w^4/64]\}^{1/2}}, \tag{74}$$

where $r_1$ is just the value of (68) appropriate for the first pion-excitation state. Although chiral Lagrangian arguments have been recently advanced suggesting that $r_1$ is substantially less than unity [4], sum-rule estimates for $r_1$ of order unity and larger [13] have received further support [25] from recent Laplace sum-rule fits.

We reiterate that the FESR-based inequality (73) [and the relation (69) from which it is derived] avoids any need for a narrow-resonance approximation, which would certainly be unphysical for dealing with broad subcontinuum pion-resonance states. The QCD-vacuum condensates that contribute are all lumped into the constant A (71); condensates such as $\langle \bar{q} G\cdot\sigma q\rangle$, $\langle \alpha_s (\bar{q} q)^2\rangle$, and $\langle \alpha_s G^3\rangle$ do not generate any $O(m_q^2)$ contributions to $F_1^L$, as has already been discussed in Sections III and IV. Even dimension-8 gluonic condensate contributions can be expected to be suppressed relative to those of $\langle \alpha_s G^2\rangle$ by the dimensional arguments presented at the end of Section IV.

In Table I, we tabulate the quark-mass lower bound $\mu(w)$ for values of $s_0$ ranging from 1 $GeV^2$ to 4 $GeV^2$. We also tabulate the same function *in the absence of instanton contributions* [i.e., with $G(w) = 0$] in order to demonstrate the key role instantons play in obtaining a lighter and phenomenologically consistent quark mass over the entire range of $s_0$ considered. When the contribution of instantons is absent (Column 4 of Table I), we find that $\mu(w)$ decreases from 9.1 MeV by a factor of four as $s_0$ increases from 1 GeV$^2$ to 4 GeV$^2$. This behaviour, if taken seriously, would not only suggest via (73) a rather large quark mass (~ 9 MeV), but also a very large aggregate contribution

$$\Sigma r_i \approx 15 \tag{75}$$

from subcontinuum resonance-peaks as $s_0$ increases to 4 GeV$^2$.

The instanton term G(w) in the denominator of (74) greatly ameliorates these effects. When the instanton term is included (Column 3 of Table I), we find that $\mu(w)$ decreases from 5.7 MeV by only a factor of two as $s_0$ goes from 1 GeV$^2$ to 4 GeV$^2$, suggesting via (73) a lighter (~ 6 MeV) quark mass in conjunction with a phenomenologically reasonable aggregate contribution

$$\Sigma r_i \approx 3 \tag{76}$$

from subcontinuum resonance-peaks as $s_0$ increases to 4 GeV$^2$. In view of the sparseness of such pion-resonance states [which suggests replacing $\Sigma$ $r_i$ with $r_1$], it is noteworthy that the estimate (76) is quite compatible with past and present sum rule estimates for $r_1$ [13,25].

It is best to regard the results presented in this section as essentially qualitative. We have utilized only the one-loop-order purely-perturbative contribution to the correlation function $\Pi^L$ − higher-order terms can be expected to alter the coefficient of the $w^4$-dependence in the denominator of (74).[5] The key point here, however, is that the function G(w) arising from instantons is oscillatory (64), going from positive to negative values as $s_0$ increases from 1 GeV$^2$ to 4 GeV$^2$. Moreover, G(w) is not only positive, but is also larger over the range 1 GeV$^2$ $\leq$ $s_0 \leq 1.6$ GeV$^2$ than the factor $w^4/64$ arising from perturbation theory, thereby lowering and stabilizing the quark mass (73) in a region for which there is at most only a partial contribution from the lowest subcontinuum resonance.

**Acknowledgments:**

ASD, VE and TGS are grateful for support from the Natural Sciences and Engineering Research Council of Canada. VE is also grateful for several useful discussions with V. A. Miransky. AHF and YX wish to acknowledge research funding from the Faculty of Graduate Studies of the University of Western Ontario.

---

[5] Inclusion of the renormalization group (RG) dependence of the running quark mass also lowers somewhat the size of the aggregate resonance contribution $\Sigma r_i$ as $s_0 \rightarrow 4$ GeV$^2$. Assuming $\Lambda_{QCD} = 0.2$ GeV, we find near-constancy of the RG-invariant quark mass $\hat{m}$ $\{m_q \rightarrow m_q(s_0) = \hat{m}/[\ln(\sqrt{s_0}/\Lambda_{QCD})]^{4/9}\}$ over the 1 GeV$^2$ $\leq s_0 \leq 4$ GeV$^2$ range of Table I provided $\Sigma r_i \rightarrow 2$ if instantons are included, with $\Sigma r_i \rightarrow 10$ if instantons are not included.

## Appendix A: Imaginary Parts of Correlation Functions for s > 4m²

The purely perturbative contribution to the longitudinal component $\Pi_L$ of the axial vector correlation function, defined via

$$i\int d^4x\; e^{ip\cdot x}\langle 0|Tj_{\mu 5}(x)j_{\nu 5}(0)|0\rangle$$
$$= [g_{\mu\nu} - p_\mu p_\nu/p^2]\Pi_T(p^2) + [p_\mu p_\nu/p^2]\Pi_L(p^2) \qquad \textbf{(A.1)}$$

with $j_{\mu 5}(x) = \bar{u}(x)\gamma_\mu\gamma_5\, d(x)$ can be obtained from Fig 4a:

$$[\Pi_L(p^2)]_{pert} \equiv C_{pert}(p^2)$$
$$= \frac{3m^2}{2\pi^2}\left[-\frac{2}{n-4} - \gamma_E + \ln\left(\frac{4\pi\mu^2}{m^2}\right) - I(p^2)\right], \qquad \textbf{(A.2)}$$

$$I(p^2) \equiv \int_0^1 dx\; \ln\left[1 - \frac{p^2}{m^2}x(1-x) - i|\varepsilon|\right]. \qquad \textbf{(A.3)}$$

If $p^2 < 4m^2$, the argument of the logarithm is positive definite, and the $i|\varepsilon|$ factor is irrelevant to the evaluation of the integral. It is straightforward to find that $C_{pert}$ is real provided $p^2 < 4m^2$, and that

$$I(p^2) = -2 + \sqrt{1 - \frac{4m^2}{p^2}}\ \ln\left[\frac{1 + \sqrt{1 - \frac{4m^2}{p^2}}}{\sqrt{1 - \frac{4m^2}{p^2}} - 1}\right], \quad p^2 < 0 \qquad \textbf{(A.4)}$$

$$I(p^2) = -2 + 2\sqrt{\frac{4m^2}{p^2} - 1}\ \tan^{-1}\left[\left(\frac{4m^2}{p^2} - 1\right)^{-1/2}\right], \quad 0 < p^2 < 4m^2 \qquad \textbf{(A.5)}$$

One easily finds from either expression that $\lim_{p^2 \to 0} I(p^2) = 0$, and from the latter expression that $\lim_{p^2 \to (4m^2)^-} I(p^2) = -2$. The results (A.2) and (A.4) are consistent with $C_{pert}$ as calculated in ref. [15]. Utilizing the notation of that reference, one sees that

$$X(v) \equiv \frac{1}{v} \ln\left[\frac{v+1}{v-1}\right] = \frac{I(p^2) + 2}{v^2} \qquad \textbf{(A.6)}$$

with

$$v = \sqrt{1 - \frac{4m^2}{p^2}}\ . \qquad \textbf{(A.7)}$$

The relationship (A.6) between $X(v)$ and $I(p^2)$, the latter quantity defined via the integral (A.3), can be utilized to determine the imaginary part of $X(v)$ when $p^2 > 4m^2$, as shown below.

If $p^2 > 4m^2$, the argument of the logarithm in the integrand of (A.3) can be factorized and integrated by parts as follows:

$$I(p^2) = \ln\left(\frac{p^2}{m^2}\right) + \int_0^1 dx\ \ln(x - \tau_+ - i\varepsilon'')$$

$$+ \int_0^1 dx\ \ln(x - \tau_- + i\varepsilon'')$$

$$= -2 - (\tau_+ + i\varepsilon'') \int_0^1 \frac{dx}{x - \tau_+ - i\varepsilon''}$$

$$- (\tau_- - i\varepsilon'') \int_0^1 \frac{dx}{x - \tau_- + i\varepsilon''} \qquad \textbf{(A.8)}$$

with

$$\tau_\pm \equiv \frac{1 \pm \sqrt{1 - \frac{4m^2}{p^2}}}{2} , \qquad \textbf{(A.9)}$$

and with

$$\varepsilon'' = \frac{p^2 \,|\, \varepsilon \,|}{m^2(\tau_+ - \tau_-)} . \qquad \textbf{(A.10)}$$

Note that if $p^2 > 4m^2$, then $0 < \tau_- < \tau_+ < 1$, and, hence, that $\varepsilon'' > 0$. Consequently, the pole in (A.8) at $\tau_+ + i\varepsilon''$ is above the real x axis, and the pole at $\tau_- - i\varepsilon''$ is below the real x axis, permitting the equivalent contours of Fig. 5 to run below $\tau_+$ and above $\tau_-$. Using the contours of Fig 5, with $C_+$ and $C_-$ assumed to be semicircles of radius $\delta$ about $x = \tau_+$ and $\tau_-$, respectively, one finds that [$C_\pm$: $z = \tau_\pm + \delta e^{i\theta_\pm}$; range of $\theta_+$: $\pi \rightarrow 2\pi$; range of $\theta_-$: $\pi \rightarrow 0$]

$$\int_0^1 \frac{dx}{x - \tau_+ - i\varepsilon''} = \operatorname*{Lim}_{\delta\to 0}\left[\int_0^{\tau_- - \delta} \frac{dx}{x-\tau_+} + \int_{\tau_- + \delta}^{1} \frac{dx}{x-\tau_+} + \int_{C_-} \frac{dz}{z-\tau_+}\right]$$

$$= \ln\frac{\tau_-}{\tau_+} + i\pi \;, \qquad \textbf{(A.11)}$$

$$\int_0^1 \frac{dx}{x-\tau_- + i\varepsilon''} = \ln\frac{\tau_+}{\tau_-} - i\pi \;. \qquad \textbf{(A.12)}$$

Substituting (A.11) and (A.12) into (A.8), one finds for $p^2 > 4m^2$ that

$$I(p^2) = -2 + v\left[\ln\left(\frac{1+v}{1-v}\right) - i\pi\right]. \qquad \textbf{(A.13)}$$

Using the relationship $X = (I+2)/v^2$, we then see that $X$ develops *a negative* imaginary part:

$$\operatorname{Im} X = -\frac{\pi}{v} \;, \quad p^2 > 4m^2 \;. \qquad \textbf{(A.14)}$$

## Appendix B: Evaluation of the Gluon Condensate Contribution to $F_0^L$

The "heavy-quark" (h.q.) gluon condensate contribution to $\Pi_L$, as defined in (A.1), is obtained from Appendix B.3 of ref [15] as the sum of coefficients $\left[C_{1G^2}\right]_{h.q.}$ and $\left[C_{2G^2}\right]_{h.q.}$ for the axial-vector current correlation function $\left[s \equiv p^2, \;\; v = \sqrt{1-4m^2/s}\right]$:

$$\left[\Pi_L(p^2)\right]_{G^2} = \left(C_{1G^2} + C_{2G^2}\right)_{h.q.} < G^2 > \ , \qquad \textbf{(B.1)}$$

$$\left[C_{1G^2}\right]_{h.q.} = \frac{\alpha}{48\pi s v^2}\left[3(1-v^2)^2 X(v) - 6(1+v^2)\right] \ , \qquad \textbf{(B.2)}$$

$$\left[C_{2G^2}\right]_{h.q.} = \frac{\alpha}{96\pi s v^4}\left[3(1-v^2)^2\ (1+v^2)\ X(v) - 2(3-2v^2+3v^4)\right] \ , \qquad \textbf{(B.3)}$$

$$\left[C_{1G^2} + C_{2G^2}\right]_{h.q.} \equiv \alpha E_{G^2} = \alpha E_{pole} + \alpha C_x\ X(v) \ , \qquad \textbf{(B.4)}$$

$$\alpha E_{pole} = -\frac{\alpha}{96\pi}\left[\frac{18}{s} + \frac{14}{s-4m^2} + \frac{24m^2}{(s-4m^2)^2}\right], \qquad \textbf{(B.5)}$$

$$\alpha C_x = \frac{\alpha}{2\pi}\ m^4\left[\frac{1}{s^3 v^4} + \frac{3}{s^3 v^2}\right]. \qquad \textbf{(B.6)}$$

We have extracted a factor of $\alpha$ so that $E_{G^2}$ as defined in (B.4) is consistent with $E_{G^2}$ as defined in Sec. 3.

The gluon condensate contribution to the finite energy sum rules

$$F_0^L = \frac{1}{2\pi i}\oint_{C(s_0)} \Pi_L(s)\ ds \ , \qquad \textbf{(B.7)}$$

$$F_1^L = \frac{1}{2\pi i} \int_{C(s_0)} \Pi_L\ (s)\ s\ ds\ , \tag{B.8}$$

can be obtained via (46) and (50) from direct evaluation of the integrals

$$G_0 \equiv \int_{C(s_0)} E_{G^2}\ ds\ , \tag{B.9}$$

$$G_1 \equiv \int_{C(s_0)} E_{G^2}\ s\ ds\ , \tag{B.10}$$

with the contour $C(s_0)$ distorted as in Fig. 6 to encompass any pole singularities of $E_{GG}$ at $s = 0$ or $4m^2$ as well as the branch singularity for $s > 4m^2$. Using Eq. (A.14), one finds that

$$\begin{aligned} G_0 = & -2i\pi \int_{4m^2+\varepsilon}^{s_0} \frac{C_x}{v}\ ds \\ & + \int_{C_0} E_{pole}\ ds + \int_{C_{4m^2}} E_{pole}\ ds \\ & + \int_{C_0} C_x\ X(v)\ ds + \int_{C_{4m^2}} C_x\ X(v)\ ds\ , \end{aligned} \tag{B.11}$$

where the contours $C_0$ and $C_{4m^2}$ are clockwise circles of radius $\varepsilon$ about $s = 0$ and $s = 4m^2$, respectively. We see from (B.5) that

$$\int_{C_0} E_{pole}\, ds = \frac{3i}{8} , \qquad \textbf{(B.12)}$$

$$\int_{C_{4m^2}} E_{pole}\, ds = \frac{7i}{24} . \qquad \textbf{(B.13)}$$

The remaining three integrals in (B.11) are evaluated as follows. Using the expression for $C_x$ in (B.6), we find that

$$-2i\pi \int_{4m^2+\varepsilon}^{s_0} \frac{C_x}{v}\, ds = -i\, m^4 \left[ I_1 + 3I_2 \right] , \qquad \textbf{(B.14)}$$

where

$$I_1 = \int_{4m^2+\varepsilon}^{s_0} \frac{1}{s^3 v^5}\, ds = \int_{4m^2+\varepsilon}^{s_0} s^{-1/2}(s-4m^2)^{-5/2}\, ds , \qquad \textbf{(B.15)}$$

$$I_2 = \int_{4m^2+\varepsilon}^{s_0} \frac{1}{s^3 v^3}\, ds = \int_{4m^2+\varepsilon}^{s_0} s^{-3/2}(s-4m^2)^{-3/2}\, ds . \qquad \textbf{(B.16)}$$

Both integrals can be evaluated via the trigonometric substitution $s = 4m^2 \sec^2\theta$. One then finds that

$$I_1 = \frac{1}{8m^4} \int_{\theta_L}^{\theta_u} \frac{\cos^3\theta}{\sin^4\theta}\, d\theta = -\frac{1}{24m^4} \left[ \frac{1}{\sin^3\theta} \right] \Bigg|_{\theta_L}^{\theta_u} + \frac{1}{8m^4} \left[ \frac{1}{\sin\theta} \right] \Bigg|_{\theta_L}^{\theta_u} , \qquad \textbf{(B.17)}$$

$$I_2 = \frac{1}{8m^4} \int_{\theta_L}^{\theta_u} \frac{\cos^3\theta}{\sin^2\theta}\, d\theta$$

$$= -\frac{1}{8m^4} \left[ \frac{1}{\sin\theta} + \sin\theta \right] \Bigg|_{\theta_L}^{\theta_u} , \qquad \textbf{(B.18)}$$

where, using the parameterization of (A.7), we find that

$$\theta_u = \sec^{-1}\left( \frac{s_0^{1/2}}{2m} \right) ; \quad \sin\, \theta_u = \sqrt{1 - 4m^2/s_0} \equiv v_0 , \qquad \textbf{(B.19)}$$

and that

$$\theta_L = \sec^{-1}\left( \frac{(4m^2+\varepsilon)^{1/2}}{2m} \right) ; \quad \sin\theta_L = \left( \frac{\varepsilon}{4m^2+\varepsilon} \right)^{1/2} . \qquad \textbf{(B.20)}$$

Substituting (B.19) and (B.20) into (B.17) and (B.18), we find from (B.14) that

$$-2i\pi \int_{4m^2+\varepsilon}^{s_0} \frac{C_x}{v}\, ds = -\frac{i}{8} \left[ -\frac{1}{3v_0^3} - \frac{2}{v_0} - 3v_0 \right] - \frac{im^3}{3\varepsilon^{3/2}} - \frac{5im}{8\varepsilon^{1/2}} + O(\varepsilon^{1/2}) . \qquad \textbf{(B.21)}$$

The integral around the origin is straightforward to obtain from (B.6) and (A.6). The integrand

$$C_x \; X(v) \;=\; \frac{1}{16\pi}\left[2+I(s)\right]\left[\frac{-3/2}{s-4m^2} + \frac{6m^2}{(s-4m^2)^2} + \frac{8m^4}{(s-4m^2)^3} + \frac{3}{2s}\right] \qquad \textbf{(B.22)}$$

has a simple pole at $s = 0$ because $I(0) = 0$ [as noted below (A.5)]:

$$\int_{C_0} C_x \; X(v) \; ds \;=\; -\frac{3i}{8} \; . \qquad \textbf{(B.23)}$$

Note that (B.23) *exactly cancels* (B.12), indicating that the origin can be excised from the contour of Fig. 6.

This cancellation is not peculiar to the channel we are in. We have verified (Appendix D) that an identical cancellation occurs in the scalar, vector, and transverse-axial channels between the contributions of explicit $s = 0$ poles in $E_{G^2}$ [as in (B.12)] and the integrals of $C_x X(v)$ portions of $E_{G^2}$ around $C_0$ [as in (B.23)]. Thus the quantum-field-theoretical singularities in $G_0$ and $G_1$ all occur for $s \geq 4m^2$ on the real s-axis for *all* of the above-mentioned channels.

The divergence as $\varepsilon \to 0$ in (B.21) is cancelled exactly by the integration of $C_x \; X(v)$, as a given in (B.22) over the contour $C_{4m^2}$ around $s = 4m^2$, a cancellation which also occurs in the other three channels mentioned above. This cancellation is most easily seen by continuing the expression (A.5) to complex values of $s$ in the vicinity of $s = 4m^2$:

$$I(s) + 2 = 2\left[(4m^2 - s)/s\right]^{1/2} \tan^{-1}\left(\frac{s}{4m^2 - s}\right)^{1/2}$$

$$= \pi\left(\frac{4m^2-s}{s}\right)^{1/2} - 2\left(\frac{4m^2-s}{s}\right) + \frac{2}{3}\left(\frac{4m^2-s}{s}\right)^2 + \ldots \qquad \textbf{(B.24)}$$

On the contour $C_{4m^2}$, $s = 4m^2 + \varepsilon e^{i\theta}$ with a clockwise rotation of $\theta$ from $2\pi$ to $0$. When $s > 4m^2$, the correct (negative) sign of the imaginary part $\left[2i \text{ Im}\{I(s)\} \equiv I(s+i|\delta|) - I(s-i|\delta|)\right]$ is obtained by requiring that

$$(4m^2 - s)^{1/2} = -i\varepsilon^{1/2} e^{i\theta/2}, \qquad \textbf{(B.25)}$$

as

$$2i \text{ Im}\{I(s)\} = \lim_{\theta\to 0}\left[\pi\left(\frac{4m^2-s}{s}\right)^{1/2}\right] - \lim_{\theta\to 2\pi}\left[\pi\left(\frac{4m^2-s}{s}\right)^{1/2}\right] \qquad \textbf{(B.26)}$$

with $s = s(\theta) = 4m^2 + |\varepsilon| e^{i\theta}$. Upon substitution of (B.24) into (B.22) one finds that

$$\int_{C_{4m^2}} C_x X(v)\, ds = -\frac{7i}{24}$$

$$+ \frac{1}{16}\left[\frac{3}{2}\int_{C_{4m^2}} s^{-1/2}(4m^2-s)^{-1/2}\, ds\right.$$

$$+ 6m^2 \int_{C_{4m^2}} s^{-1/2}(4m^2-s)^{-3/2}\, ds$$

$$- 8m^4 \int_{C_{4m^2}} s^{-1/2}(4m^2-s)^{-5/2}\, ds$$

$$\left. + \frac{3}{2}\int_{C_{4m^2}} s^{-3/2}(4m^2-s)^{1/2}\, ds\right]. \qquad \textbf{(B.27)}$$

The factor $-7i/24$ is just $-2\pi i$ times the aggregate residue at $s = 4m^2$ obtained from multiplication of (B.24)'s integer powers of $(4m^2 - s)$ into (B.22). This pole contribution *explicitly cancels* the pole contribution (B.13). The remaining integrals in (B.27) result from multiplying the leading $\pi\left[(4m^2 - s)/s\right]^{1/2}$ term of (B.24) into (B.22). These integrals are easily evaluated around the clockwise contour $C_{4m^2}$ via (B.25):

$$\int_{C_{4m^2}} (4m^2 - s)^{-1/2}\ s^{-1/2}\ ds \ = \ O(\varepsilon^{1/2})\ , \qquad \textbf{(B.28)}$$

$$\int_{C_{4m^2}} (4m^2 - s)^{-3/2}\ s^{-1/2}\ ds \ = \ \frac{2i}{m\varepsilon^{1/2}} + O(\varepsilon^{1/2})\ , \qquad \textbf{(B.29)}$$

$$\int_{C_{4m^2}} (4m^2 - s)^{-5/2}\ s^{-1/2}\ ds \ = \ -\frac{2i}{3m\varepsilon^{3/2}} + \frac{i}{4m^3\varepsilon^{1/2}} + O(\varepsilon^{1/2})\ , \qquad \textbf{(B.30)}$$

$$\int_{C_{4m^2}} (4m^2 - s)^{1/2}\ s^{-3/2}\ ds \ = \ O(\varepsilon^{3/2})\ . \qquad \textbf{(B.31)}$$

Substituting (B.28 - B.31) into (B.27) we find that

$$\int_{C_{4m^2}} C_x\ X(v)\ ds \ = \ -\frac{7i}{24} + \frac{5im}{8\varepsilon^{1/2}} + \frac{im^3}{3\varepsilon^{3/2}} + O(\varepsilon^{1/2}), \qquad \textbf{(B.32)}$$

explicitly cancelling the divergencies in (B.21). Since all the $s=0$ and $s=4m^2$ pole terms contributing to $G_0$ have also been shown to cancel, we find that $G_0$ is equal to the upper-bound contribution of the first integral on the right-hand side of (B.11):

$$G_0 = -2i\pi \int^{s_0} \frac{C_x}{v}\, ds = \frac{i}{8}\left[\frac{1}{3v_0^3} + \frac{2}{v_0} + 3v_0\right];$$

$$v_0 \equiv \sqrt{1-4m^2/s_0}\ . \qquad \textbf{(B.33)}$$

To obtain the full contribution of $<\alpha_s G^2>$ to the $F_0$ sum rule, we substitute Eq. (46) from the text into (50), utilizing the results (B.33) in conjunction with Eqs. (30) and (34) from the text:

$$\left[F_0^L(s_0)\right]_{<\alpha_s G^2>} = <\alpha_s G^2> \left\{\frac{1}{16\pi}\left[\frac{1}{3v_0^3} + \frac{2}{v_0} + 3v_0\right] - \frac{1}{3\pi}\right\}$$

$$= <\alpha_s G^2> \left[\frac{m^4}{\pi s_0^2} + \frac{14m^6}{3\pi s_0^3} \ldots\right]. \qquad \textbf{(B.34)}$$

**Appendix C: The Gluon Condensate Contribution to $F_1^L$**

Consider first the integral $G_1$ (B.10), which can be evaluated via the following integrals arising from the distortion of $C(s_0)$ indicated in Fig 6:

$$G_1 = -2i\pi \int_{4m^2+\varepsilon}^{s_0} \frac{C_x}{v}\, s\, ds + \int_{C_0} E_{pole}\, s\, ds$$

$$+ \int_{C_{4m^2}} E_{pole}\, s\, ds + \int_{C_0} C_x\, X(v)\, s\, ds$$

$$+ \int_{C_{4m^2}} C_x\, X(v)\, s\, ds\ . \qquad \textbf{(C.1)}$$

One sees from (B.5) that

$$\int_{C_0} E_{pole}\ s\ ds = 0\ , \tag{C.2}$$

$$\int_{C_{4m^2}} E_{pole}\ s\ ds = \frac{+5im^2}{3}\ . \tag{C.3}$$

Using the expression for $C_x$ in (B.6), we find that

$$-2i\pi \int_{4m^2+\varepsilon}^{s_0} \frac{C_x}{v}\ s\ ds = -im^4\left[I_3 + 3I_4\right]\ , \tag{C.4}$$

where the integrals $I_3$ and $I_4$ are evaluated using (B.19-20), as in the previous section:

$$\begin{aligned} I_3 &\equiv \int_{4m^2+\varepsilon}^{s_0} \frac{1}{s^2 v^5}\ ds = -\frac{1}{6m^2}\left[\left.\frac{1}{\sin^3\theta}\right|_{\theta_L}^{\theta_u}\right] \\ &= -\frac{1}{6m^2 v_0^3} + \frac{4m}{3\varepsilon^{3/2}} + \frac{1}{2m\varepsilon^{1/2}} + O(\varepsilon^{1/2})\ , \end{aligned} \tag{C.5}$$

$$\begin{aligned} I_4 &= \int_{4m^2+\varepsilon}^{s_0} \frac{1}{s^2 v^3}\ ds = \frac{1}{2m^2}\left.\left[-\frac{1}{\sin\theta}\right]\right|_{\theta_L}^{\theta_u} \\ &= -\frac{1}{2m^2 v_0} + \frac{1}{m\varepsilon^{1/2}} + O(\varepsilon^{1/2})\ . \end{aligned} \tag{C.6}$$

Substituting (C.5) and (C.6) into (C.4) we find that

$$-2i\pi \int_{4m^2+\varepsilon}^{s_0} \frac{C_x}{v}\ s\ ds = im^2 \left[ \frac{1}{6v_0^3} + \frac{3}{2v_0} \right] - \frac{4im^5}{3\varepsilon^{3/2}} - \frac{7im^3}{2\varepsilon^{1/2}} \ . \qquad \textbf{(C.7)}$$

Using (B.22), we find that $C_x\ X(v)\ s$ has no poles at $s = 0$ [note that $2 + I(0) = 2$], in which case

$$\int_{C_0} C_x\ X(v)\ s\ ds = 0 \ . \qquad \textbf{(C.8)}$$

Once again, we note that the origin can be excised entirely from the contour of Fig 6. We have verified explicitly that integrals (C.2) and (C.8) are zero in the scalar, vector and transverse axial channels as well (see Appendix D).

As in the previous section, the divergence in (C.7) as $\varepsilon \to 0$ is exactly cancelled by integration of $C_x\ X(v)\ s$ around the contour $C_{4m^2}$. From (B.22) we find that

$$C_x\ X(v)\ s = \frac{2}{\pi} \left[2 + I(s)\right] \left[ \frac{m^4}{(s-4m^2)^2} + \frac{m^6}{(s-4m^2)^3} \right]. \qquad \textbf{(C.9)}$$

If we substitute (B.24) into (C.9) and integrate around $C_{4m^2}$, we easily separate a pure-pole contribution from an $\varepsilon$-dependent contribution involving half-integral powers of $(4m^2 - s)$:

$$\begin{aligned} \int_{C_{4m^2}} C_x\ X(v)\ s\ ds &= -\frac{5im^2}{3} + 2m^4 \int_{C_{4m^2}} (4m^2-s)^{-3/2}\ s^{-1/2}\ ds \\ &\quad - 2m^6 \int_{C_{4m^2}} (4m^2-s)^{-5/2}\ s^{-1/2}\ ds \\ &= -\frac{5im^2}{3} + \frac{7im^3}{2\varepsilon^{1/2}} + \frac{4im^5}{3\varepsilon^{3/2}} + O(\varepsilon^{1/2}) \ . \qquad \textbf{(C.10)} \end{aligned}$$

The final line of (C.10) is obtained through use of (B.29) and (B.30). Not only are the $\varepsilon$-

dependent terms in (C.7) cancelled by the final line of (C.10), but the pure-pole contribution (C.3) also cancels against the pole term in (C.10). Thus we find, as in Appendix B, that $G_1$ is equal to the upper-bound contribution of the first integral on the right-hand side of (C.1):

$$G_1 = -2i\pi \int^{s_0} \frac{C_x\, s}{v}\, ds = im^2 \left[ \frac{1}{6v_0^3} + \frac{3}{2v_0} \right]. \qquad \textbf{(C.11)}$$

To obtain the full contribution of $<\alpha_s G^2>$ to the $F_1$ sum rule, we again substitute Eq. (46) of the text into (51), utilizing the results (C.11) in conjunction with Eqs. (30) and (34) from the text:

$$\left[F_1^L(s_0)\right]_{<\alpha_s G^2>} = \frac{m^2}{2\pi} <\alpha_s G^2> \left\{ \left[ \frac{1}{6v_0^3} + \frac{3}{2v_0} \right] - \frac{2}{3} \right\}$$

$$= \frac{m^2}{2\pi} <\alpha_s G^2> \left\{ 1 + \frac{4m^2}{s_0} + \frac{14m^4}{s_0^2} + \frac{160m^6}{3s_0^3} \ldots \right\}. \qquad \textbf{(C.12)}$$

## Appendix D: Gluon Condensate Contributions to $F_{0,1}$ in Other Channels

Utilizing the notation and conventions of Appendices B and C, we list the following results obtained from scalar, vector and the transverse component of the axial-vector correlation functions:

*Scalar Channel*

From Appendix B.1 of ref 15, we have

$$\left[C_{G^2}\right]_{h.q.} \equiv \alpha E_{G^2} = \alpha\left(E_{pole} + C_x \, X(v)\right), \quad \textbf{(D.1)}$$

$$E_{pole} = \frac{(3 - v^2)}{16\pi s v^2}, \quad \textbf{(D.2)}$$

$$C_x = -\frac{(1 - v^2)(3 + v^2)}{32\pi s v^2} \, . \quad \textbf{(D.3)}$$

We find that

$$\int_{C_0} E_{pole} \, ds = \frac{i}{8} \, , \quad \textbf{(D.4)}$$

$$\int_{C_{4m^2}} E_{pole} \, ds = -\frac{3i}{8} \, , \quad \textbf{(D.5)}$$

$$-2i\pi \int_{4m^2+\varepsilon}^{s_0} \frac{C_x}{v} \, ds = \frac{i}{8}\left[-\frac{3}{v_0} + v_0 + \frac{6m}{\sqrt{\varepsilon}}\right], \quad \textbf{(D.6)}$$

$$\int_{C_0} C_x \, X(v) \, ds = -\frac{i}{8} \, , \quad \textbf{(D.7)}$$

$$\int_{C_{4m^2}} C_x \, X(v) \, ds = \frac{3i}{8}\left(1 - \frac{2m}{\sqrt{\varepsilon}}\right). \quad \textbf{(D.8)}$$

Summing (D.3-8) we obtain

$$G_0 = \int_{C(s_0)} E_{G^2}\, ds = \frac{i}{8}\left[-\frac{3}{v_0} + v_0\right]. \quad \textbf{(D.9)}$$

We also find from Appendix B.1 of ref. 15 that

$$\int_{C(s_0)} C_{\bar{q}q}\, ds = 6i\pi \ , \quad \textbf{(D.10)}$$

$$\int_{C(s_0)} C_M\, ds = 0 \ , \quad \textbf{(D.11)}$$

which implies via (46) and (60) that

$$\left[F_0(s_0)\right]_{<\alpha_s G^2>} = \frac{1}{16\pi}\left[-\frac{3}{v_0} + v_0 + 4\right] <\alpha_s G^2> \ . \quad \textbf{(D.12)}$$

Unlike the case of $F_0$, the FESR $F_1$ requires the use of (46) to eliminate a logarithmic mass singularity in $G_1$, obtained by summing the following five integrals:

$$\int_{C_0} E_{pole} s\, ds = 0 \ , \quad \textbf{(D.13)}$$

$$\int_{C_{4m^2}} E_{pole} s\, ds = -\frac{3im^2}{2} \quad \textbf{(D.14)}$$

$$-2i\pi \int_{4m^2+\varepsilon}^{s_0} \frac{C_x}{v}\, s\, ds = \frac{im^2}{2}\left[-\frac{3}{v_0} + 2\ell n(1-v_0^2) + \frac{6m}{\sqrt{\varepsilon}}\right], \quad \textbf{(D.15)}$$

$$\int_{C_0} C_x \; X(v) \; s \; ds \; = \; 0 \; , \tag{D.16}$$

$$\int_{C_{4m^2}} C_x \; X(v) \; s \; ds \; = \; \frac{3im^2}{2} \; - \; \frac{3m^3}{\sqrt{\varepsilon}} \; . \tag{D.17}$$

We then find that

$$G_1 \; = \; \int_{C(s_0)} E_{G^2} \; s \; ds \; = \; \frac{im^2}{2}\left[-\frac{3}{v_0} \; + \; 2\ell n\left(\frac{4m^2}{s_0}\right)\right], \tag{D.18}$$

which is not analytic in $m$ at $m = 0$. However the results

$$\int_{C(s_0)} C_{\overline{q}q} \; s \; ds \; = \; 4im^2\pi \; , \tag{D.19}$$

$$\int_{C(s_0)} C_M \; s \; ds \; = \; -2im\pi \; , \tag{D.20}$$

used in conjunction with (46) and (51) eliminates the quark-mass from the logarithm:

$$\left[F_1(s_0)\right]_{<\alpha_s G^2>} \; = \; \frac{m^2}{2\pi}\left[-\frac{3}{2v_0} \; + \; \frac{1}{3} \; - \; \ell n\left(\frac{s_0}{4\mu^2}\right)\right]<\alpha_s G^2> \; . \tag{D.21}$$

*Transverse Axial Channel*

From Appendix B.3 of ref. 15, we have

$$\left[C_{1G^2}\right]_{h.q.} \equiv \alpha\ E_{G^2} = \alpha\left(E_{pole} + C_x\ X(v)\right)\ , \quad \textbf{(D.22)}$$

$$E_{pole} = -\frac{(1+v^2)}{8\pi s v^2}\ , \quad \textbf{(D.23)}$$

$$C_x = \frac{(1-v^2)^2}{16\pi s v^2}\ . \quad \textbf{(D.24)}$$

We then find that

$$\int_{C_0} E_{pole}\ ds = \frac{i}{4}\ , \quad \textbf{(D.25)}$$

$$\int_{C_{4m^2}} E_{pole}\ ds = \frac{i}{4}\ , \quad \textbf{(D.26)}$$

$$-2i\pi \int_{4m^2+\varepsilon}^{s_0} \frac{C_x}{v}\ ds = \frac{i}{4}\left[\frac{1}{v_0} - v_0 - \frac{2m}{\sqrt{\varepsilon}}\right], \quad \textbf{(D.27)}$$

$$\int_{C_0} C_x\ X(v)\ ds = -\frac{i}{4}\ , \quad \textbf{(D.28)}$$

$$\int_{C_{4m^2}} C_x\ X(v)\ ds = -\frac{i}{4} + \frac{im}{2\sqrt{\varepsilon}}\ . \quad \textbf{(D.29)}$$

As before, the contour-radius singularity as $\varepsilon \to 0$ cancels between (D.27) and (D.29):

$$G_0 = \int_{C(s_0)} E_{G^2}\, ds = \frac{i}{4}\left[\frac{1}{v_0} - v_0\right]. \qquad \textbf{(D.30)}$$

Since in this channel, one finds that [15]

$$\int_{C(s_0)} C_{\bar{q}q}\, ds = -4\pi i\ , \qquad \textbf{(D.31)}$$

$$\int_{C(s_0)} C_M\, ds = 0\ , \qquad \textbf{(D.32)}$$

we find via (46) and (60) that

$$\left[F_0(s_0)\right]_{<\alpha_s G^2>} = \frac{1}{8\pi}\left[\frac{1}{v_0} - v_0 - \frac{4}{3}\right]<\alpha_s G^2>\ . \qquad \textbf{(D.33)}$$

Corresponding results for $F_1$ are listed below:

$$\int_{C_0} E_{pole}\, s\, ds = 0\ , \qquad \textbf{(D.34)}$$

$$\int_{C_{4m^2}} E_{pole}\, s\, ds = im^2\ , \qquad \textbf{(D.35)}$$

$$-2i\pi \int_{4m^2+\varepsilon}^{s_0} \frac{C_x}{v}\ s\ ds = im^2\left(\frac{1}{v_0} - \frac{2m}{\sqrt{\varepsilon}}\right), \tag{D.36}$$

$$\int_{C_0} C_x\ X(v)\ s\ ds = 0\ , \tag{D.37}$$

$$\int_{C_{4m^2}} C_x\ X(v)\ s\ ds = -im^2 + \frac{2im^3}{\sqrt{\varepsilon}}\ , \tag{D.38}$$

$$\int_{C(s_0)} C_{\bar{q}q}\ s\ ds = -\frac{8i\pi m^2}{3}\ , \tag{D.39}$$

$$\int_{C(s_0)} C_M\ s\ ds = 0\ , \tag{D.40}$$

$$\left[F_1(s_0)\right]_{<\alpha_s G^2>} = \left(\frac{m^2}{2\pi v_0} - \frac{m^2}{9\pi}\right)<\alpha_s G^2>\ . \tag{D.41}$$

*Vector Channel*

From eq. (II.19) of ref. 15, we find that

$$E_{pole} = -\frac{(3-2v^2+3v^4)}{48\pi s v^4} \,, \tag{D.42}$$

$$C_x = \frac{(1-v^2)^2(1+v^2)}{32\pi s v^4} \,. \tag{D.43}$$

We then find $F_0$ (D.51) from $G_0$, the sum of (D.44-48), in conjunction with (46) and (D.49-50):

$$\int_{C_0} E_{pole}\, ds = \frac{i}{8} \,, \tag{D.44}$$

$$\int_{C_{4m^2}} E_{pole}\, ds = \frac{i}{24} \,, \tag{D.45}$$

$$-2i\pi \int_{4m^2+\varepsilon}^{s_0} \frac{C_x}{v}\, ds = \frac{i}{8}\left[v_0 + \frac{1}{3v_0^3} - \frac{8m^3}{3\varepsilon^{3/2}} - \frac{m}{\varepsilon^{1/2}}\right], \tag{D.46}$$

$$\int_{C_0} C_x\, X(v)\, ds = -\frac{i}{8} \,, \tag{D.47}$$

$$\int_{C_{4m^2}} C_x\, X(v)\, ds = -\frac{i}{24} + \frac{im^3}{3\varepsilon^{3/2}} + \frac{im}{8\varepsilon^{1/2}} \,, \tag{D.48}$$

$$\int_{C(s_0)} C_{\bar{q}q}\, ds = -4i\pi \,, \tag{D.49}$$

$$\int_{C(s_0)} C_M \, ds = 0 \ , \qquad \textbf{(D.50)}$$

$$\left[F_0(s_0)\right]_{<\alpha_s G^2>} = \frac{1}{16\pi}\left(v_0 + \frac{1}{3v_0^3} - \frac{8}{3}\right)<\alpha_s G^2> \ . \qquad \textbf{(D.51)}$$

Corresponding results for $F_1$ are listed below:

$$\int_{C_0} E_{pole} \, s \, ds = 0 \ , \qquad \textbf{(D.52)}$$

$$\int_{C_{4m^2}} E_{pole} \, s \, ds = \frac{2im^2}{3} \ , \qquad \textbf{(D.53)}$$

$$-2i\pi \int_{4m^2+\varepsilon}^{s_0} \frac{C_x}{v} \, s \, ds = \frac{im^2}{2}\left[\frac{1}{v_0} + \frac{1}{3v_0^3} - \frac{8m^3}{3\varepsilon^{3/2}} - \frac{3m}{\varepsilon^{1/2}}\right], \qquad \textbf{(D.54)}$$

$$\int_{C_0} C_x \, X(v) \, s \, ds = 0 \ , \qquad \textbf{(D.55)}$$

$$\int_{C_{4m^2}} C_x\ X(v)\ s\ ds = -\frac{2im^2}{3} + \frac{4im^5}{3\varepsilon^{3/2}} + \frac{3im^3}{2\varepsilon^{1/2}}\ , \qquad \textbf{(D.56)}$$

$$\int_{C(s_0)} C_{\bar{q}q}\ s\ ds = -\frac{16}{3}\ i\pi m^2\ , \qquad \textbf{(D.57)}$$

$$\int_{C(s_0)} C_M\ s\ ds = 0\ , \qquad \textbf{(D.58)}$$

$$\left[F_1(s_0)\right]_{<\alpha_s G^2>} = \frac{m^2}{4\pi}\left[\frac{1}{v_0} + \frac{1}{3v_0^3} - \frac{8}{9}\right]<\alpha_s G^2>\ . \qquad \textbf{(D.59)}$$

**Figure Captions:**

Figure 1: a) The contour $C(s_0)$.

b) Distortion of $C(s_0)$ to enclose the positive real s-axis.

Figure 2: The Breit-Wigner resonance shape $y(s) = m\Gamma/[(s-m^2)^2+m^2\Gamma^2]$ expressed as a sum of symmetric square pulses. For an infinitesimally thin pulse at a given value of y, the pulse will extend from $s = m^2 - [m\Gamma/y - m^2\Gamma^2]^{1/2}$ to $s = m^2 + [m\Gamma/y - m^2\Gamma^2]^{1/2}$. If there are n such pulses, the $j^{th}$ pulse is at $y = j/(nm\Gamma)$.

Figure 3: a) Leading $\langle\overline{q}q\rangle$ contribution to current correlation functions.

b) Typical leading $\langle\overline{q}G\cdot\sigma q\rangle$ contribution to current correlation functions.

c) Typical leading $\langle\alpha_s(\overline{q}q)^2\rangle$ contribution to current correlation functions.

Figure 4: a) Leading purely-perturbative contribution to current correlation functions.

b) Typical leading $\langle\alpha_s G^2\rangle$ contributions to current correlation functions.

c) Typical leading $\langle G^3\rangle$ contributions to current correlation functions.

Figure 5: Distortion of the integration contour along the real s-axis consistent with the location of the $\tau_\pm$ singularities in the complex s-plane.

Figure 6: Distortion of the $C(s_0)$ contour [Fig. (1a)] for $\langle\alpha_s G^2\rangle$ contributions to $F_{0,1}$ sum rules.

| w | $s_0$ in $GeV^2$ [Eq.(70)] | $\mu(w)$ in MeV [Eq.(74), with G(w) given by (64)] | $\mu(w)$ in MeV [Eq.(74), with G(w)=0] |
|---|---|---|---|
| 3.3 | 0.98 | 5.7 | 9.1 |
| 3.5 | 1.10 | 5.2 | 8.2 |
| 3.7 | 1.23 | 4.8 | 7.4 |
| 3.9 | 1.37 | 4.5 | 6.7 |
| 4.1 | 1.51 | 4.2 | 6.1 |
| 4.3 | 1.66 | 4.0 | 5.5 |
| 4.5 | 1.82 | 3.8 | 5.1 |
| 4.7 | 1.99 | 3.7 | 4.7 |
| 4.9 | 2.16 | 3.5 | 4.3 |
| 5.1 | 2.34 | 3.4 | 4.0 |
| 5.3 | 2.53 | 3.4 | 3.7 |
| 5.5 | 2.72 | 3.3 | 3.4 |
| 5.7 | 2.92 | 3.2 | 3.2 |
| 5.9 | 3.13 | 3.1 | 3.0 |
| 6.1 | 3.34 | 3.05 | 2.8 |
| 6.3 | 3.57 | 3.0 | 2.6 |
| 6.5 | 3.80 | 2.9 | 2.5 |
| 6.7 | 4.04 | 2.8 | 2.3 |

**Table I:** Behaviour of $\mu(w)$ with increasing $s_0$ in the presence (Column 3) and in the absence (Column 4) of direct single instanton contributions to the $F_1^L$ finite energy sum rule.

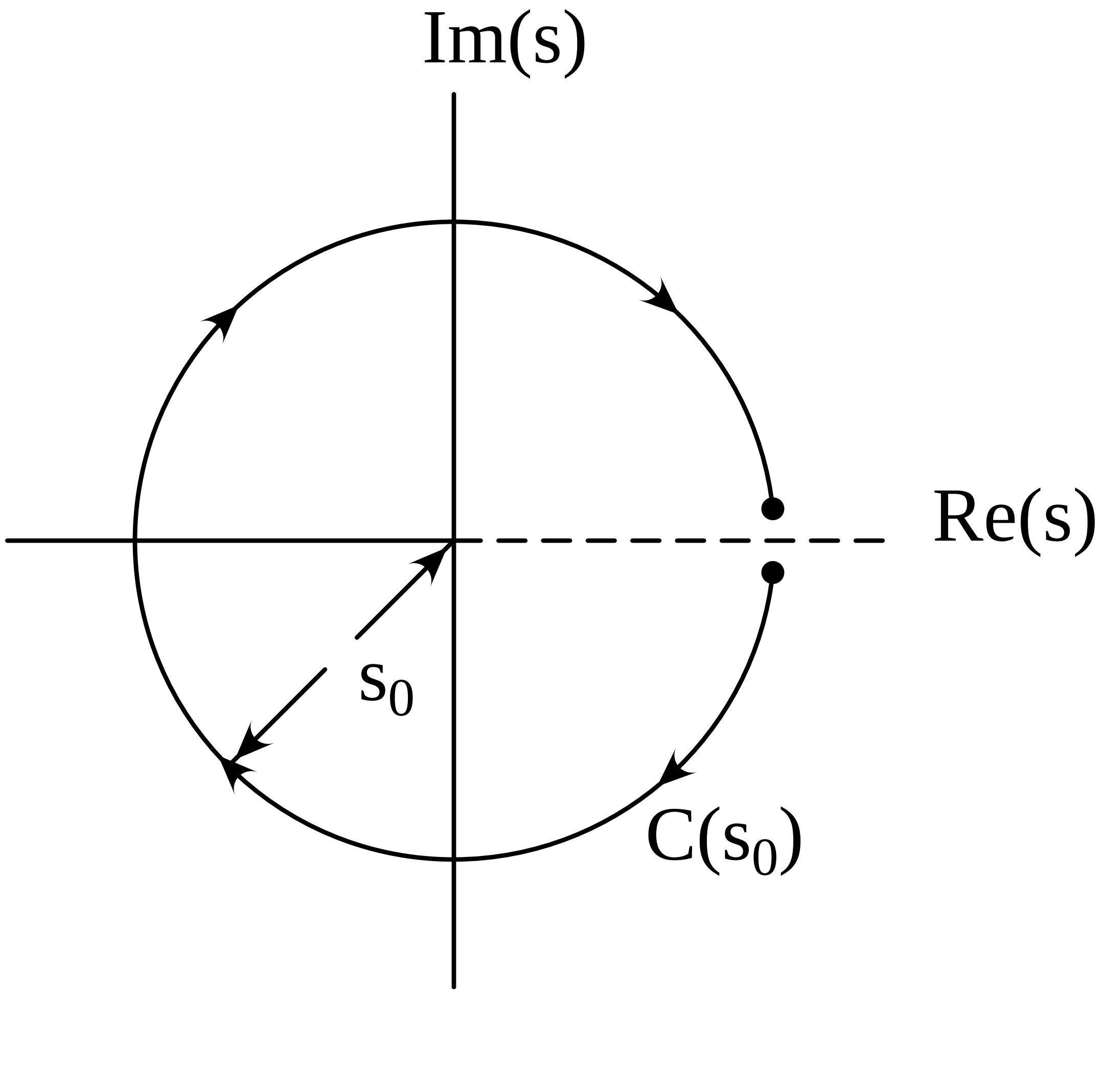

Fig. 1a

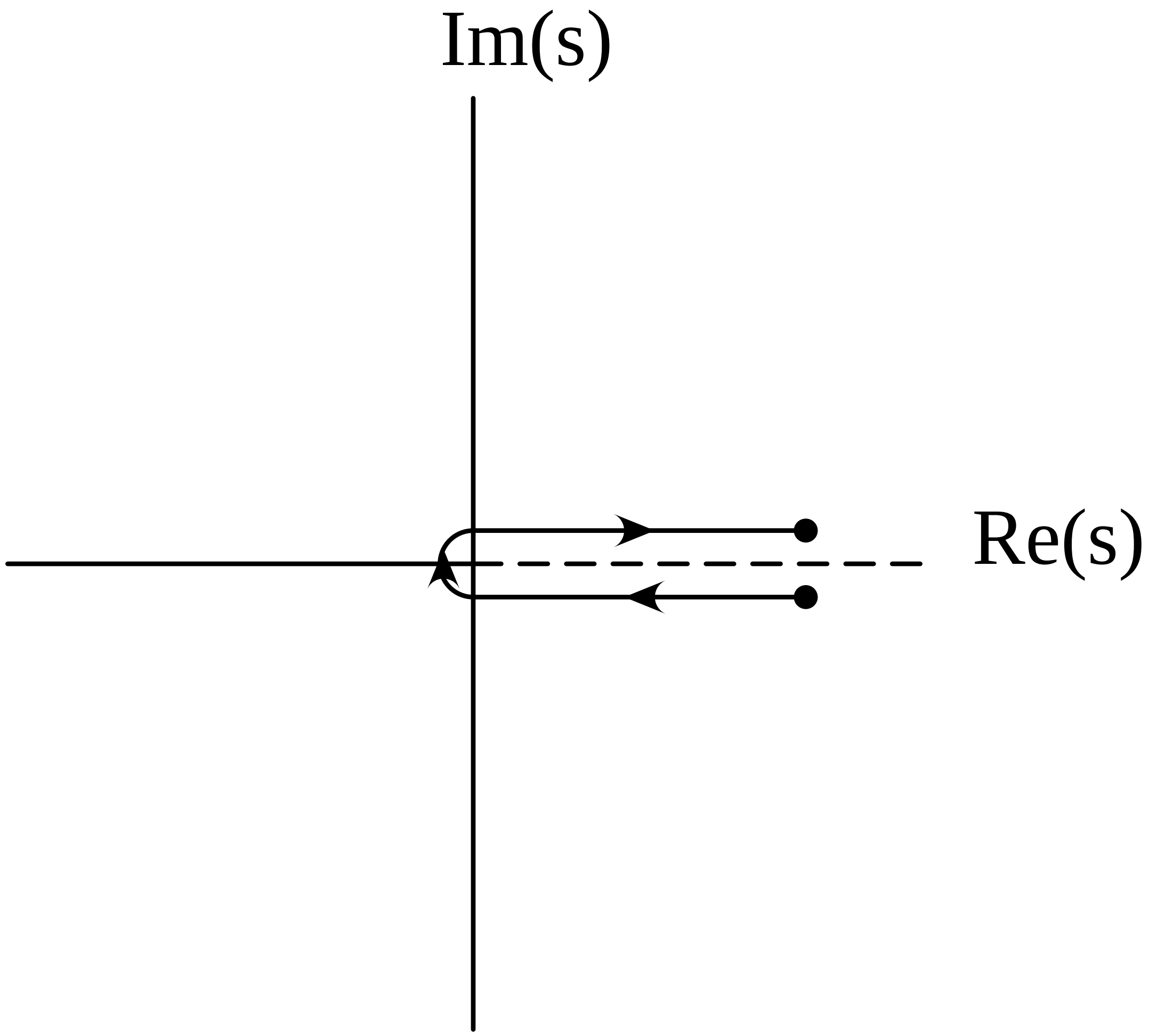

Fig. 1b

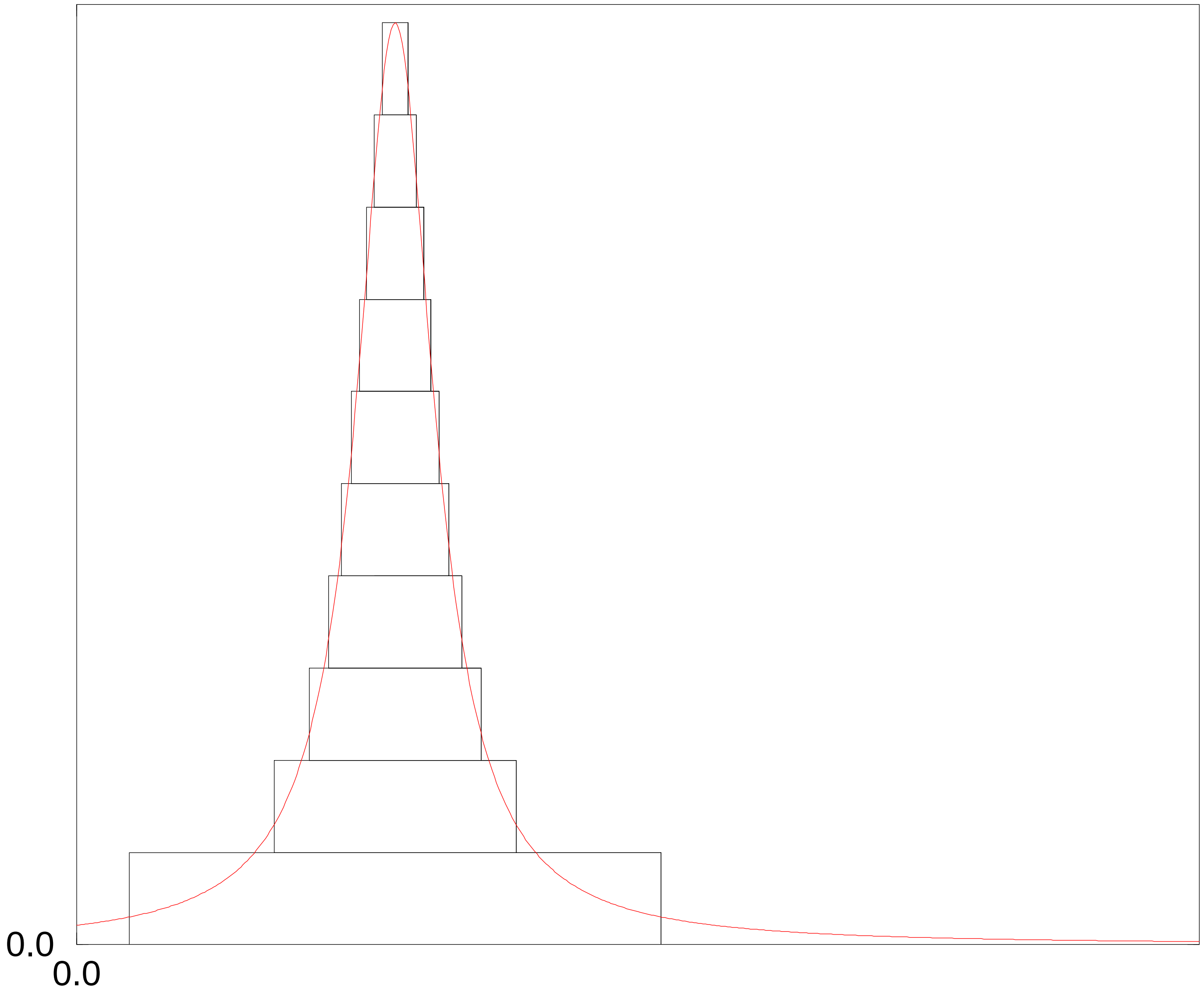
0.0
0.0

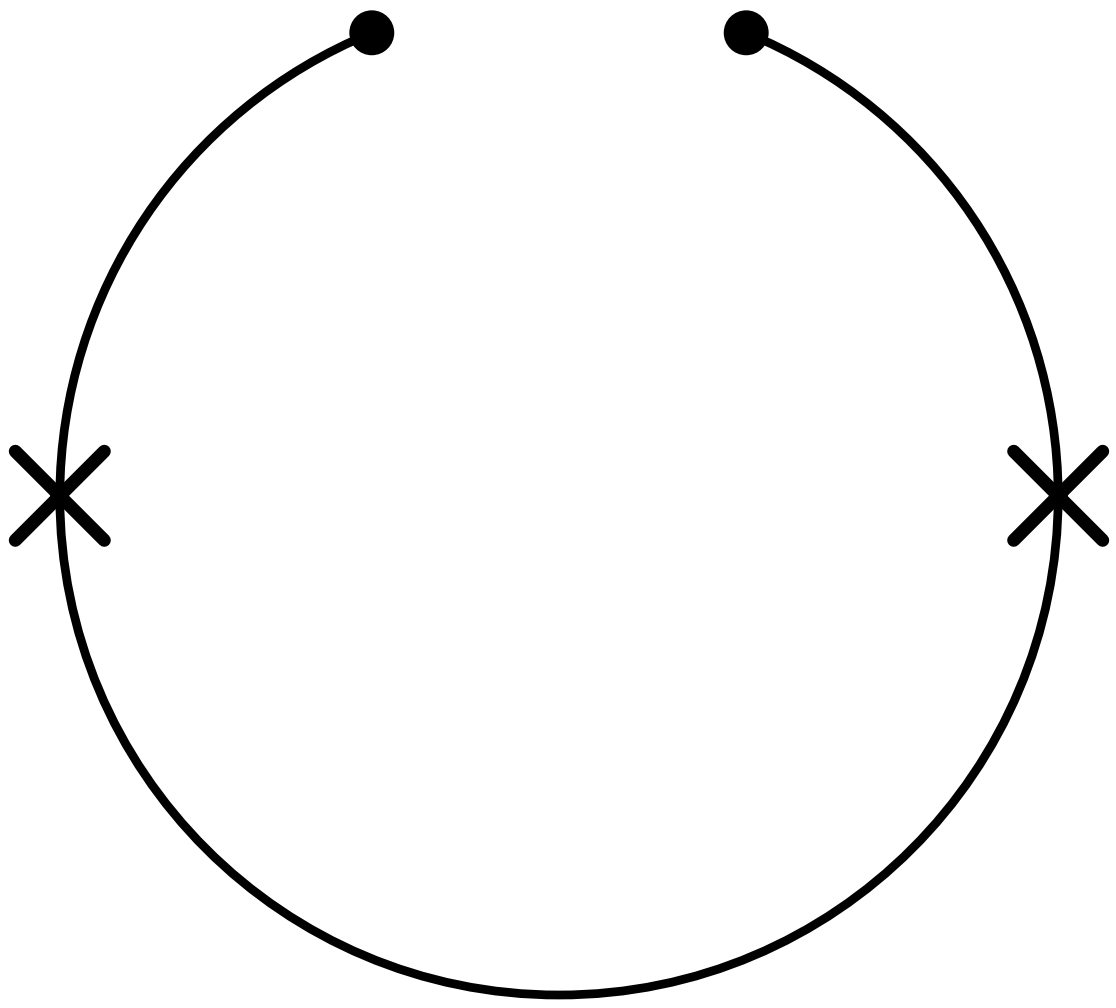

Fig. 3a

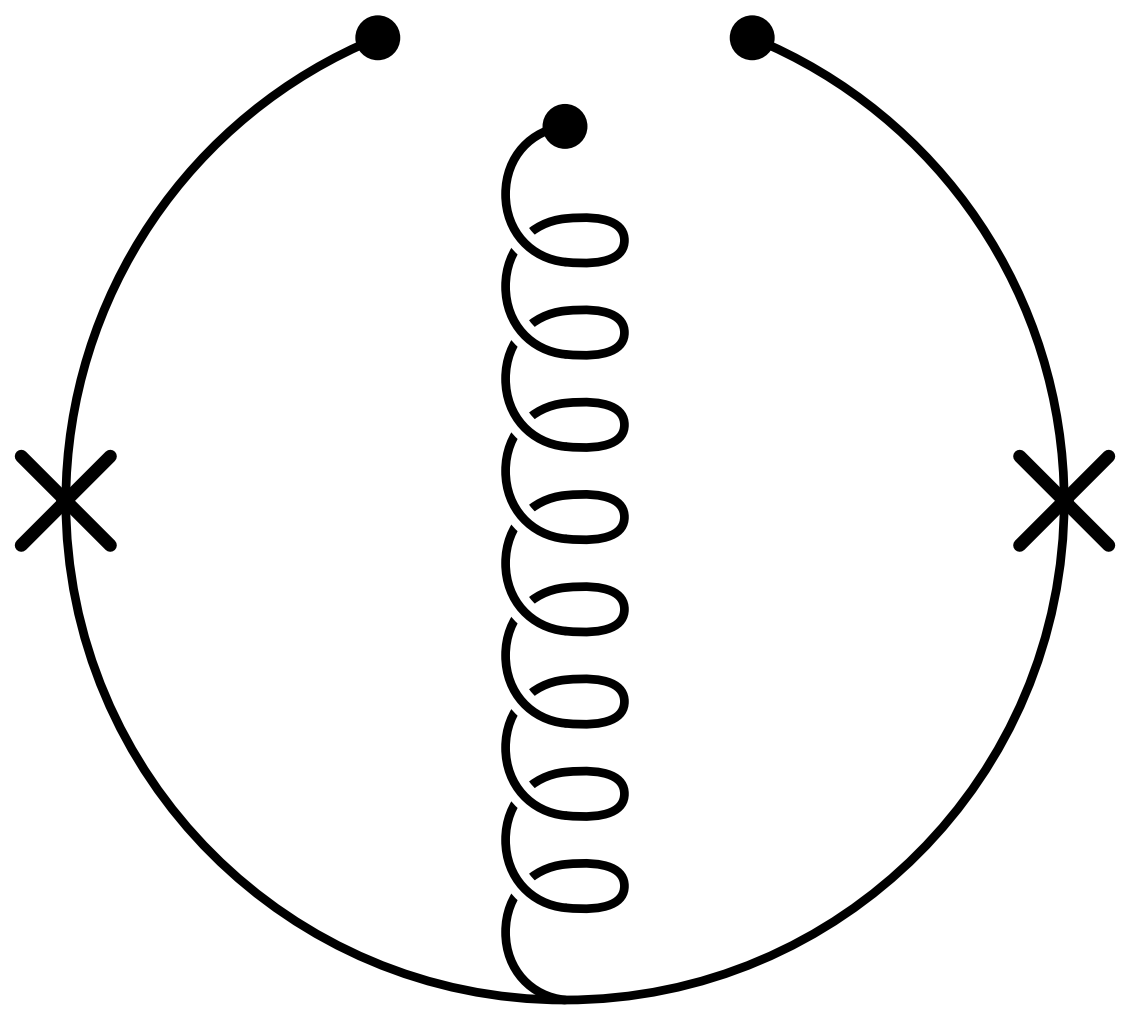

Fig. 3b

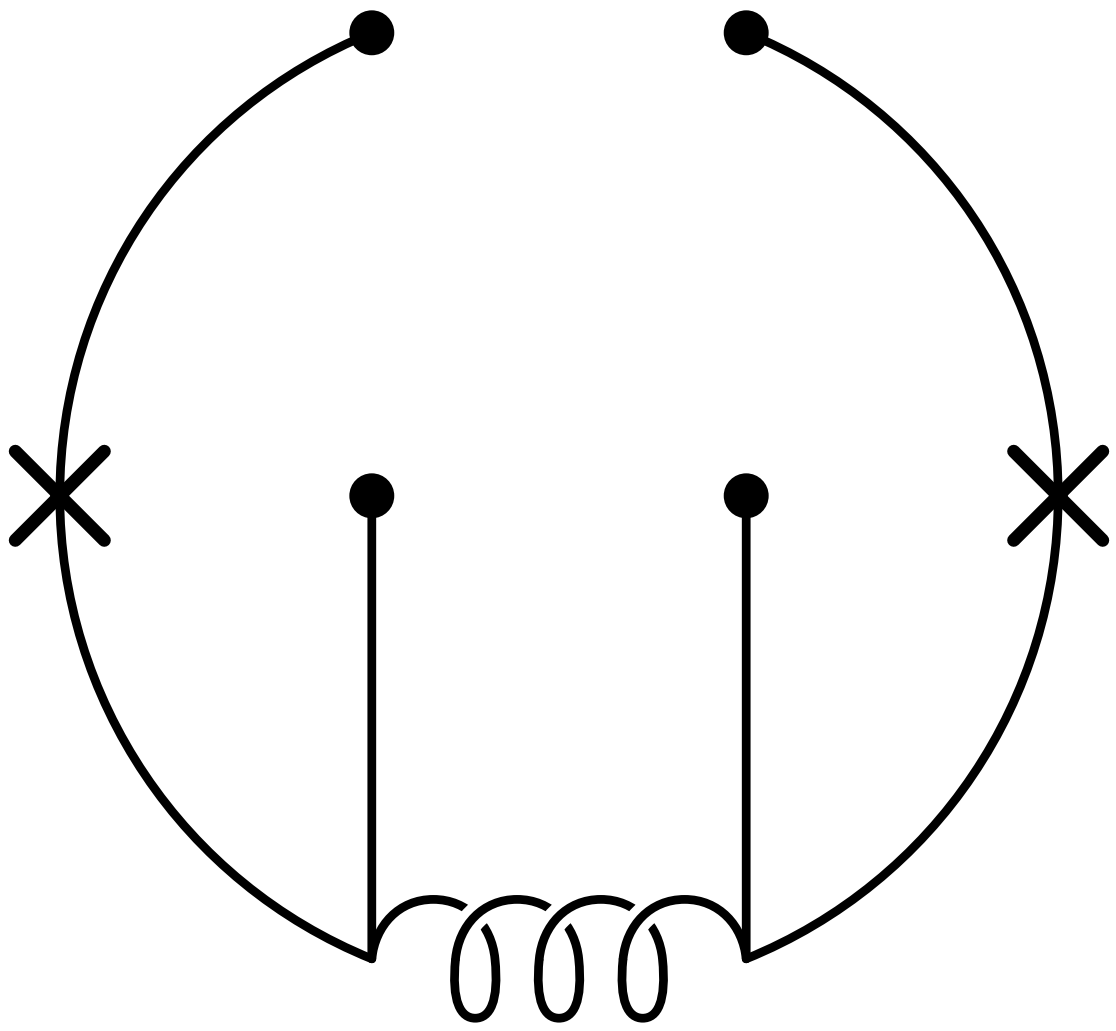

Fig. 3c

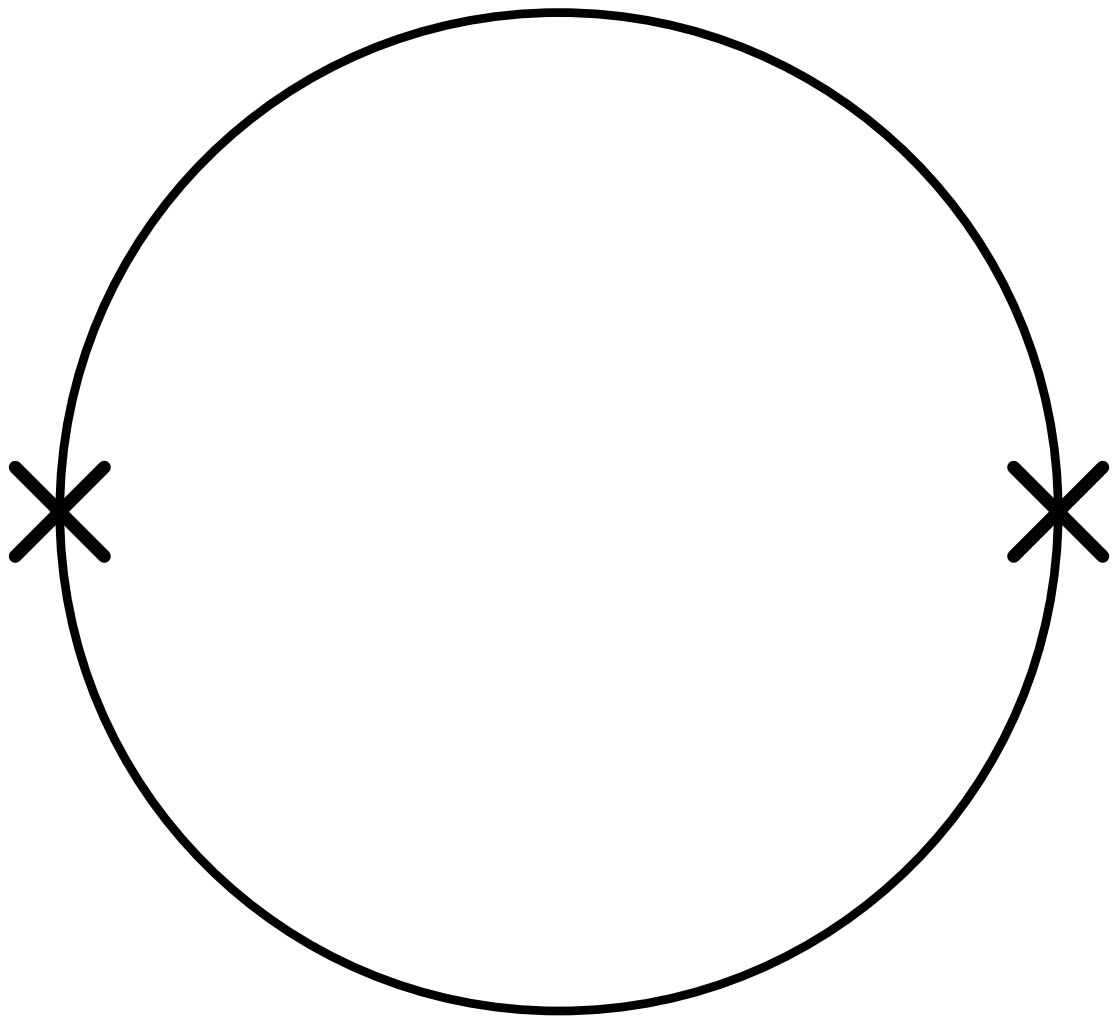

Fig. 4a

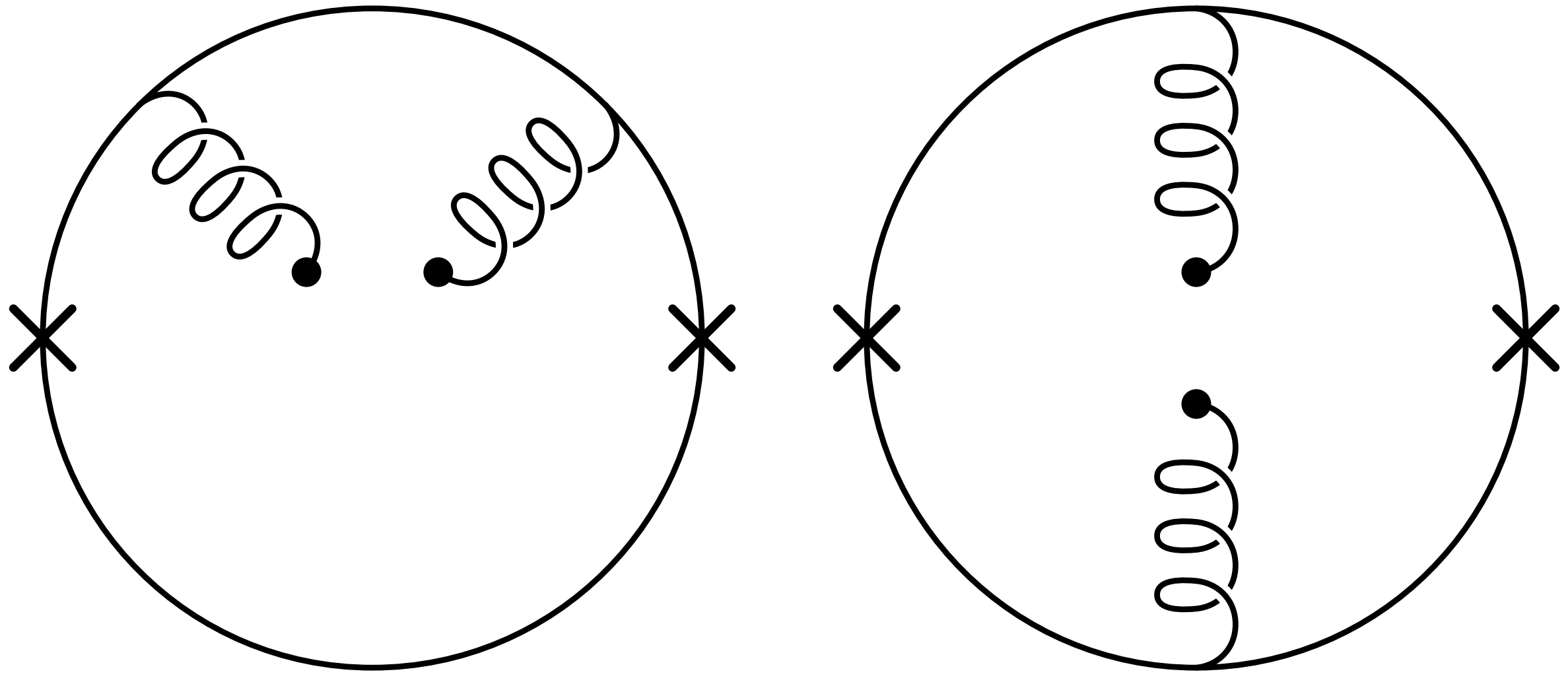

Fig. 4b

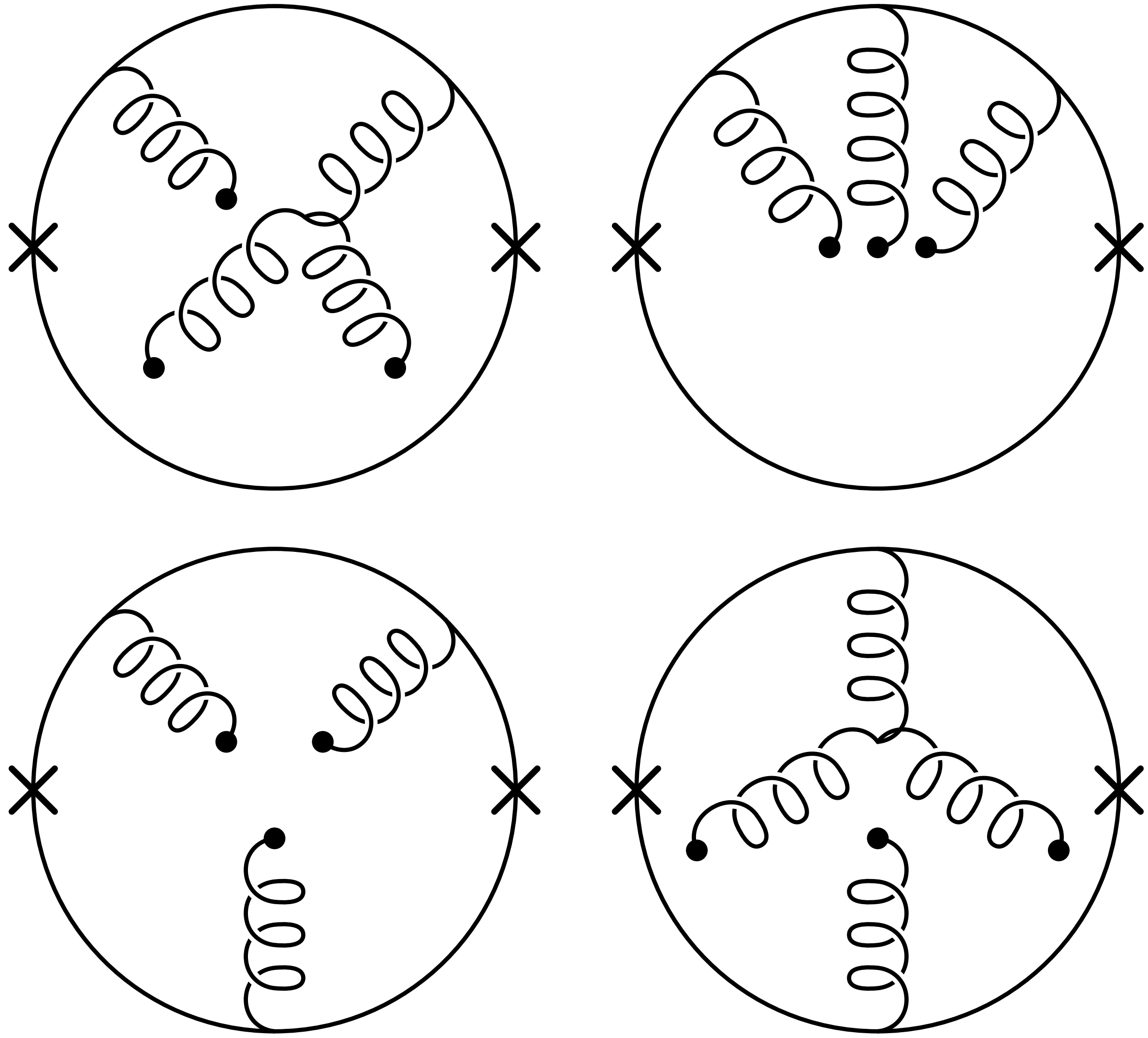

Fig. 4c

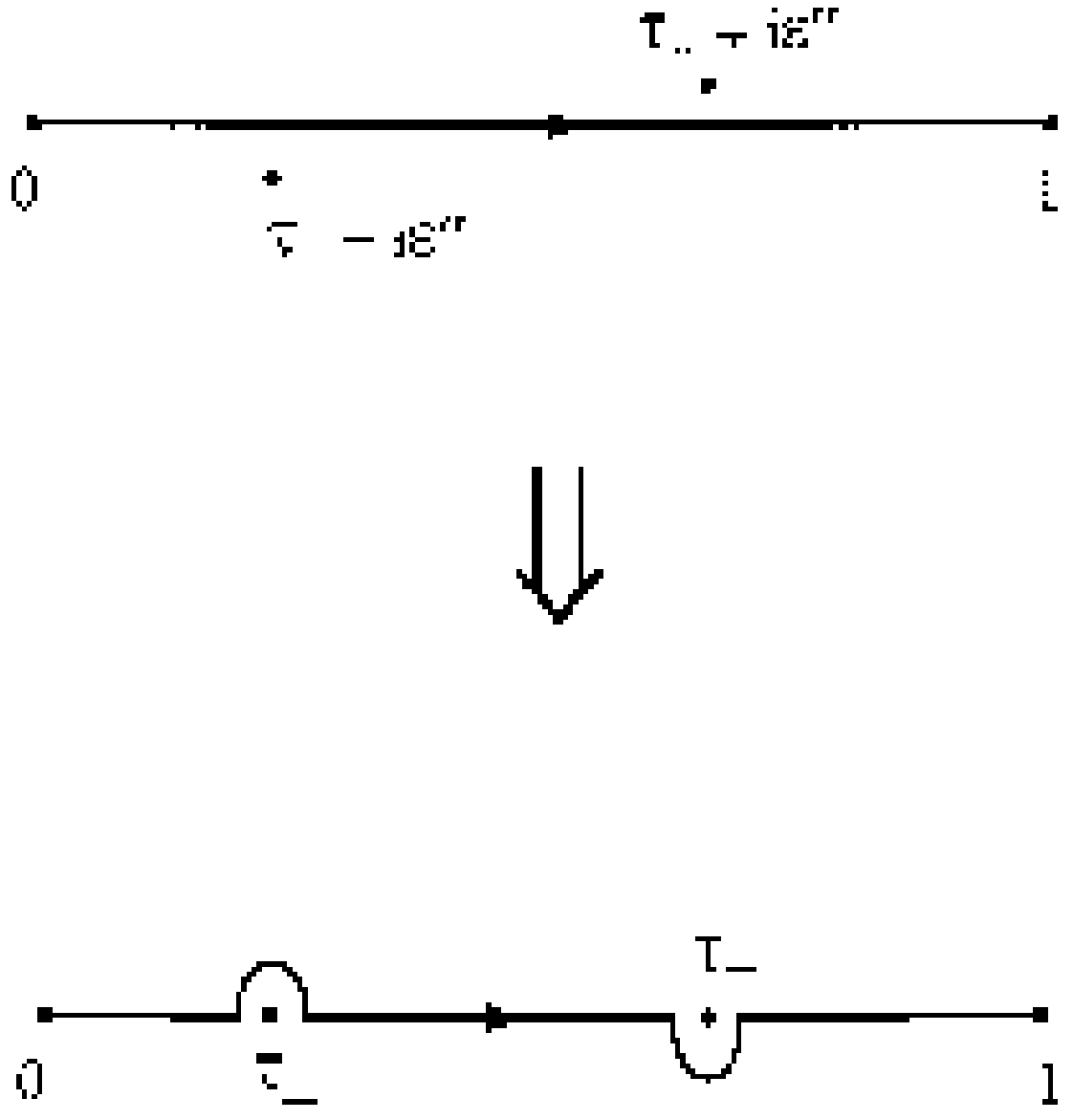
0
0
1

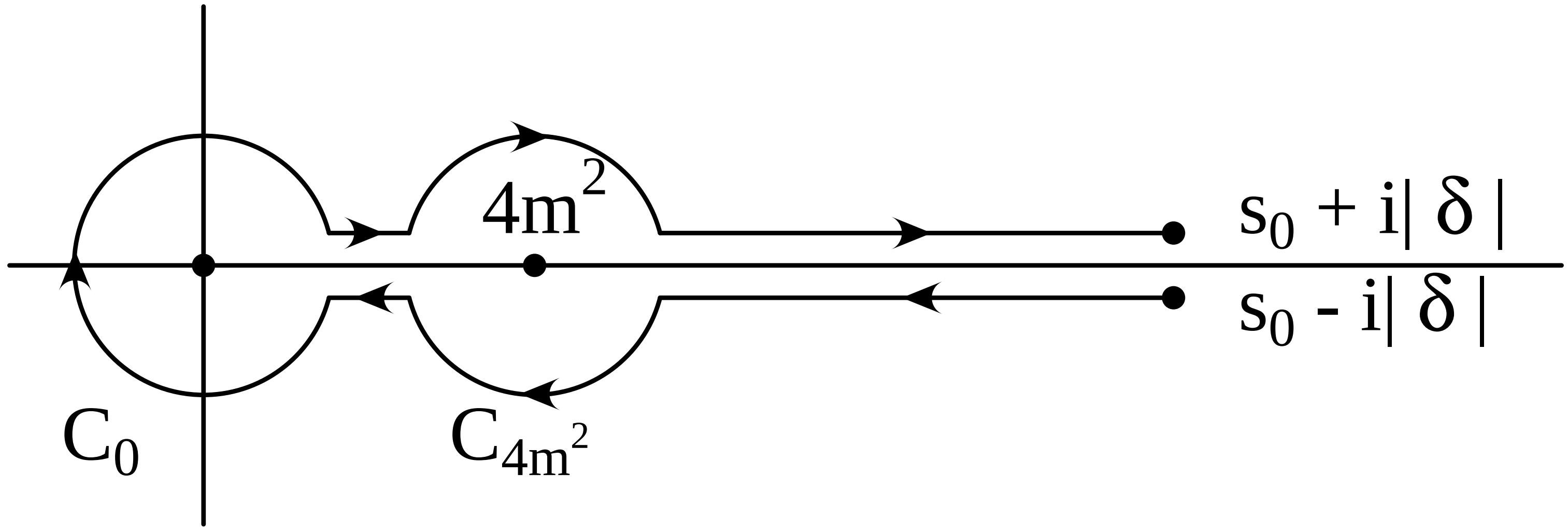

Fig. 6